\documentstyle[preprint,floats,aps,epsf,epsfig,prb]{revtex}
\newlength{\bxwidth}\bxwidth=2.5 truein
\newcommand\om { \omega}
\newcommand\eps { \epsilon}
\newcommand\frm[1]{\epsfig{file=#1,width=\bxwidth}}
\newlength{\fight}
\newcommand{\fg}[3]
{\begin{figure}[tb]\epsfysize=\fight 
\centerline{\epsfbox{#1}}\vskip 0.4truein
\caption{{#2}}\label{#3}\end{figure}}
\newcommand\ltdash{\raise-1.8pt\hbox{$\scriptscriptstyle |$}}
\newcommand \beq  {\begin{equation}}
\newcommand \eeq  {\end{equation}}
\newcommand \bea {\begin{eqnarray} }
\newcommand \eea {\end{eqnarray}}
\newcommand \up{\uparrow}
\newcommand \dw{\downarrow}

\newcommand \mat[1]{\left[\matrix{#1}\right]}
\newcommand \D{\partial_{\tau}}
\newcommand \rarrow{\rightarrow}
\newcommand \dg{^{\dagger}}
\newcommand \si { \sigma}
\newcommand \ra { \rangle}
\newcommand\la{\langle}
\newcommand \Ls { {\cal L}_{ susy}}
\newcommand \Ps { \Psi_{\si}}

\newcommand\bs[1]{b_{#1}}
\newcommand\fs[1]{f_{#1}}
\begin{document}
\draft
%***********    This is for two columns *******************************
%\twocolumn[\hsize\textwidth\columnwidth\hsize\csname @twocolumnfalse\endcsname
%********************************
\title{
Supersymmetric spin operators }
\author{  P.
Coleman
$^{1}$ , C. P\'epin $^2$ and A. M. Tsvelik
$^{2}$}
\address{$^1$ Center for Materials Theory,
Department of Physics and Astronomy, 
Rutgers University, Piscataway, NJ 08854, USA.
}
\address{$^2$ Department of Physics, 
Oxford University, 
1 Keble Road, 
Oxford OX1 3P, UK.}
\maketitle
\date{\today}
\maketitle
%\widetext
\begin{abstract}
We develop a
supersymmetric representation of spin operators
which unifies the Schwinger
and Abrikosov representations of SU(N) spin operators,
allowing a second-quantized treatment of 
representations of the SU(N) group with both
symmetric and antisymmetric character.
By applying this to the SU(N) Kondo model, we
show that it is possible to develop a controlled
treatment of both Magnetism and the Kondo effect
within a single large N expansion. 
\end{abstract}
\vskip 0.2 truein
\newpage
% ***********    This is for two columns *******************************
%\vskip2pc]
%********************************

\section{Motivation for a new spin representation}

Recent experiments on  quantum phase transitions in heavy
fermion materials have led to a debate about how
magnetism condenses out of the metallic state at
absolute zero.    Certain heavy fermion materials
can can be tuned  between the magnetic and the
paramagnetic state through the use of
pressure\cite{mathur,grosche} or chemical 
doping\cite{lohneyson,schroeder}.    The quantum
critical point which separates these two phases
is of great current interest, in part  because
materials in its vicinity  may become
fundamentally new  kinds of
metal\cite{stewart,hilbert2,steglich}. Heavy
fermion materials contain a dense lattice of
magnetic moments; conventional wisdom assumes
that the spins  of the local moments are
magnetically screened and of no importance to the
magnetic  quantum critical
point\cite{hertz,millisqcp,rosch99}. 
Recent neutron data contradict this viewpoint,
by showing that the
spin correlations at the quantum critical point are critical in time,
but local on an atomic
scale\cite{schroeder,sces,schroeder99,qimiao99}, suggesting that unscreened
local moments emerge from the metallic state at the quantum
critical point.  

If it is indeed
true that the magnetic quantum critical points involve local moment
physics, then a  new theoretical approach 
is required. 
Traditionally, heavy fermion physics
is modeled using a  Kondo lattice Hamiltonian,\cite{doniach77} describing
the interaction between a bath of conduction electrons and
an array of local moments.  
One of the well-developed theoretical methods for
approaching this model is the large $N$
expansion\cite{coleman,read,readimp,auerbach,millis}, where the idea
is
to use a  generalization
of the quantum mechanical spin operators, in which the
underlying spin
rotation group is generalized  from $SU(2)$ to $SU(N)$. The  utility of this
method derives from the fact that in the limit  $N\rarrow
\infty$, it provides an
essentially exact, analytic treatment of the Kondo lattice problem.
Unfortunately the way this procedure is carried out at present,  
magnetic
interactions are suppressed as a $1/N^2$ correction,  beyond
the horizon for a controlled computation.
In this paper, we show how we can overcome this
shortcoming by the use
of a supersymmetric spin representation for local moments.  

\def\sig{\pmb{$\sigma$}}
\def\spi{\pmb{$\Gamma$}}
The theoretical description of interacting local moments poses a
fundamental problem: the Pauli Spin operator $\pmb{$S$}$ does not
satisfy a Wick's decomposition theorem, which pre-empts its use in
a Feynman diagram approach. The traditional solution is to
represent the spin in terms of either bosons, or fermions.  
In the
``Schwinger boson'' approach,\cite{auerbach88,arovas} the  spin operator 
is represented in terms of an $N$-component  boson $b_{\alpha}$; 
in the
alternative ``Abrikosov pseudo-fermion'' representation,\cite{abrikosov} the spin is
represented by an $N$ component fermion, $f_{\alpha}$, as follows

\setlength {\unitlength}{0.004\textwidth}
\begin{center}
\begin{picture}(200,100)(0,0)
\put(0,80){(a) Antisymmetric}
\put(20,60){\vector(0,1){10}}
\put(19,50){$Q$}
\put(20,40){\vector(0,-1){10}}
\put(45,30){\framebox(10,40){}}
\put(45,40){\line(1,0){10}}
\put(45,60){\line(1,0){10}}
\put(50,45){.}
\put(50,50){.}
\put(50,55){.}
\put(10,10){${\bf S}_f=f\dg_{\alpha}\spi_{\alpha \beta}f_{\beta}$}
\put(10,0){$n_f=Q$}
\put(125,80){(b) Symmetric}
\put(120,50){\framebox(60,10){}}
\put(130,50){\line(0,1){10}}
\put(170,50){\line(0,1){10}}
\put(140,55){.}
\put(150,55){.}
\put(160,55){.}
\put(147,40){$2S$}
\put(125,10){${\bf S}_b=b\dg_{\alpha}\spi_{\alpha
\beta}b_{\beta}$}
\put(125,0){$n_b=2S$}
\put(140,45){\vector(-1,0){20}} 
\put(160,45){\vector(1,0){20}} 
\end{picture} 
\end{center}
\vskip 0.1truein

\noindent Fig. 1. Young tableaux\cite{grptheory}
for (a) antisymmetric 
and (b) symmetric representations generated by the Abrikosov fermion and
Schwinger boson representations respectively. 
\vskip 0.2truein
\noindent Here,   $\spi\equiv ( \Gamma^1, \dots
\Gamma^{M})$ represents the $M= (N^2-1)/2$ independent
SU(N) generators.  
By combining $2S$ ``Schwinger bosons'' together, one 
generates a {\sl symmetric} representation of  SU(N), denoted by a
horizontal Young-Tableau with
$2S$ boxes (Fig. 1b).
Conversely, in the pseudo-fermion approach.
$Q\le N$ spin fermions are combined to 
generate an ${
\sl antisymmetric}$
representation of SU(N), denoted by a column Young-Tableau
with $Q$ boxes, (Fig. 1a).  

Most cases of physical interest 
correspond to an elementary spin, 
or a single box in the Young-Tableau. 
Unfortunately, to develop a controlled Feynman diagram expansion,
we are obliged to consider a large number of such boxes,
letting $N\rarrow \infty$ keeping either $q= Q/N$ or $m= 2S/N$ fixed.
In the process of letting $N\rarrow \infty$, some essential physics
is lost. Symmetric representations are ideal for treating magnetism,
where the ordered moment involves a highly symmetric condensation of
spin bosons, but they lose all information about the Fermi liquid fixed point.
Antisymmetric representations capture the development
of the Kondo effect and heavy fermion bands, but  magnetism is 
suppressed. 

Various authors have tried to develop alternative representations of the spin operator\cite{georges,zarand} and in this context one interesting idea is to use supersymmetry to simultaneously express the bosonic and fermionic character of local moments\cite{gan,pepin,ngai}.
%We have found a pair of supersymmetric constraints that unifies the Schwinger %boson and the Abrikosov pseudo-fermion approaches, making it possible to carry% out a controlled large-$N$ expansion that is able to deal with both magnetism% and the Kondo effect.

\section{Super Spins}

We now examine a  class of spin representations
which preserve both   symmetric and antisymmetric
correlations.   Consider the spin operator that is a sum of $n_f$ fermions
and
$n_b$ bosons, given by
\bea
{\bf S} = { \bf S} _f +{ \bf S} _b \label{sum}
\eea
By combining $Q=n_f+ n_b$ bosons and fermions together, we generate
"L-shaped" representations of SU(N). For each choice of $n_f$ and $n_b$,
we generate two irreducible representations. 
For example,  we can combine one fermion and two
bosons  as follows:

\begin{center}
\begin{picture}(200,40)(0,0)
\put(25,10){\framebox(10,10){}}
\put(27,12){f}
\put(42,12){$\otimes$}
\put(55,10){\framebox(20,10){}}\put(65,10){\line(0,1){10}}
\put(57,12){b}\put(67,12){b}
\put(82,12){$=$}
\put(100,3){\framebox(10,20){}}
\put(110.5,13){\framebox(10,10){}}
\put(100,13){\line(1,0){10}}
\put(102,5){f}\put(102,15){b}\put(112,15){b}
\put(130,12){$\oplus$}
\put(150,10){\framebox(30,10){}}
\put(160,10){\line(0,1){10}}
\put(170,10){\line(0,1){10}}
\put(152,12){f}\put(162,12){b}\put(172,12){b}
\put(95,-10){``$S=1/2$\ "}
\put(150,-10){``$S=3/2$"}
\end{picture} 
\end{center}
\vskip 0.3truein
\noindent 
For $SU(2)$, these correspond to a spin $1/2$, 
and a spin $3/2$ representation. 
To uniquely
parameterize an {\sl irreducible} representation, we need to  fix the
Cazimir 
$
{\bf S^2 } \equiv \sum_{a} S_a S_a \label{caz}
$.
Consider an L-shaped representation of SU(N) of 
width $w$, height $h$ :

\setlength {\unitlength}{0.006\textwidth}
\begin{center}
\begin{picture}(100,40)(0,0)
\multiput(20,25)(5,0){6}{\framebox(5,5){}}
\put(30,37){\vector(-1,0){10}}
\put(40,37){\vector(1,0){10}}
\put(31,35){$w$}
\multiput(20,20)(0,-5){4}{\framebox(5,5){}}
\put(13,22){\vector(0,1){8}}
\put(10,17){$h$}
\put(13,13){\vector(0,-1){8}}
\end{picture}\end{center}
\noindent If the generators of the fundamental representation are
normalized according to
$
{\rm Tr} [ \Gamma^a\Gamma^b] = \delta^{ab}$,
then the
expression for the 
Cazimir of an arbitrary irreducible
representation is\cite{zarand,okubo} 
\bea
{\bf S }^2 = \frac{Q(N^2 - Q)}{N} + \sum_{j=1,h}m_j(m_j + 1 - 2 j)
\eea
where 
$m_j$ is the number of boxes in the $j-th$ row from the top
and $Q$ is the total number of boxes.
For an L-shaped Young tableau,  $(m_1,m_2 \dots m_h)= (w,1,1\dots, 1)$, 
so that 
\bea
{\bf S }^2 &=& \frac{Q(N^2 - Q)}{N} + w(w-1) - h(h-1)
\eea
If we substitute 
$Q= w+h-1$ and $Y= h-w$ we then obtain
\bea
{\bf S }^2 =
Q(N- { Y} - Q/N).
\eea
In this way, each irreducible L-shaped representation of SU(N)
is uniquely defined by the two quantities
$(Q,{  Y})$, where ${ Y}$ can assume the values
\bea
{ Y}= -Q+1, -Q+3, \dots Q-1.\eea 
For example, if 
$Q=3$, there are three irreducible representations:

\setlength {\unitlength}{0.004\textwidth}
\begin{picture}(200,40)(0,0)
\put(30,20){${ Y}=$}\put(63,20){$2$}\put(100,20){$0$}
\put(140,20){$-2$}
\put(60,-20){\framebox(10,30){}}\put(60,0){\line(1,0){10}}
\put(60,-10){\line(1,0){10}}
\put(90,-10){\framebox(10,20){}}\put(90,0){\line(1,0){10}}
\put(100.5,0){\framebox(10,10){}}
\put(130,0){\framebox(30,10){}}\put(140,0){\line(0,1){10}}
\put(150,0){\line(0,1){10}}
\end{picture} 
\vskip 0.6truein
\noindent 

We now seek to cast both $Q$ and ${ Y} $ in an operator language. 
In terms of the boson and fermion
operators, the Cazimir can be written 
\bea
\hat {\bf S}^2 = (\hat { \bf S} _f +\hat { \bf S} _b)^2  
\eea
If we expand this expression (appendix A) by using the completeness
relation
\bea
N \Gamma^a_{\alpha \beta} \Gamma^a_{\delta\gamma} = N
\delta_{\alpha \gamma} \delta_{\beta\delta} - 
 \delta_{\alpha \beta} \delta_{\delta\gamma} \label{complete}
\eea
we are able to express the Cazimir in operator form:
\bea
\hat {\bf S}^2 =  \hat Q ( N - \hat {\cal Y}- \frac{\hat Q}{N})\label{cazimir}
\eea
where now
\bea
\hat Q= n_f+ n_b,
\eea
fixes the number of boxes and
\bea 
\hat{\cal Y}= n_f - n_b + \overbrace{\frac{1}{Q}[{\theta} , 
{\theta}\dg ]}^{\cal P},
\eea
is the operator measuring the asymmetry $h-w$ of the
representation. Here   we have introduced the operators
\bea
{\theta }\dg =f\dg
_{\beta}b_{\beta}, \qquad { \theta}  = b\dg_{\alpha}f_{\alpha}.
\eea
If we wish to study a spin system described by the
$(Q,{ Y})$ representation, then we must restrict our attention
to states $\vert \psi\ra$ in the Hilbert space which satisfy
\bea
\hat Q \vert \psi \ra = Q \vert \psi\ra\cr
\hat {\cal Y} \vert \psi \ra = { Y} \vert \psi\ra
\eea
Curiously, although this constrains the total number of bosons and
fermions, the difference $n_b-n_f$ is only partially 
constrained, reflecting the fact that bosons and fermions can
inter-convert without altering the representation. 

When we represent  spins in terms of bosons and fermions, each box
in the Young tableau is associated with a fermion or
boson, where fermions
occupy the column, bosons the row.
The corner of the tableau can contain either
a boson, or a fermion.
The operator $\theta\dg $ converts the corner box from a boson into a fermion,
whilst 
$\theta$ converts it back again.  These operators are
the generators of a ``supergroup'' $SU(1|1)$\cite{supergroups}, with the algebra
\bea
\{\theta \dg, \theta\} &=& Q.\label{comm}\eea
The spin operator commutes with these generators
\bea
[ { \bf S}_b+ { \bf S_f}, \theta]= [ { \bf S}_b+ { \bf S_f}, \theta\dg]= 0,
\eea
so that the representation is supersymmetric. 
$Q$ and
${\cal Y}$ also commute with $\theta$ and $\theta\dg$, and these are the
Cazimirs of this group. From (\ref{comm}), we see that operators 
$P^b= \frac{1}{Q}\theta \theta\dg$ and
$P^f= \frac{1}{Q}\theta\dg \theta$ satisfy $P_b+ P_f=1$: they are
the projection operators   which respectively project out 
states with ``bosons" or ``fermions" in the corner
of the Young tableau. 
In this way, we see that
${\cal P}= P_f-P_b$ is   $+1$ or $-1$, depending
on whether the state has a boson, or fermion
in the corner of the representation 
For example, the representation given by $(Q, { Y}) = (3,0)$
can be written in two ways:
\begin{center}
\begin{picture}(200,90)(0,0)
\put(130,60){\framebox(10,20){}}
\put(136,40){${\cal P} = -1$}
\put(120,30){$n_f-n_b= 1$}
\put(136,20){${ Y}=0$}
\put(115,5){ $b\dg_{\si}(f\dg_{\uparrow}f\dg_{\dw})
\vert 0 \ra
$}
\put(130,70){\line(1,0){10}}
\put(140.5,70){\framebox(10,10){}}
\put(142,72){$b$}
\multiput(132,73)(0,-10){2}{$f$}
\put(90,80){$\theta\dg$}
\put(70,75){\vector(1,0){40}}
\put(110,70){\vector(-1,0){40}}
\put(90,63){$\theta$}
\put(30,59.5){\framebox(10,10){}}
\put(26,40){${\cal P} = 1$}
\put(10,30){$n_f-n_b= -1$}
\put(26,20){${ Y}=0$}
\put(-5,5){e.g $\qquad\frac{1}{\sqrt{3}}b\dg_{\si}(b\dg_{\uparrow}f\dg_{\dw}-
b\dg_{\dw}f\dg_{\up}) \vert 0 \ra
$}
\put(40,70){\line(0,1){10}}
\put(30,70){\framebox(20,10){}}
\put(42,72){$b$}\put(32,72){$b$}
\put(32,63){$f$}
\end{picture} 
\end{center}
\vskip 0.1truein
\noindent 
The  invariance of the representation under the boson-fermion
transformation is a manifestation of the supersymmetry. 
When a boson is converted into
a fermion, the change in $n_f-n_b$ is compensated by the change in ${\cal
P}$, so that ${\cal Y}$ is invariant.

We can alternatively write the constraints in terms
of the height $\hat h = (\hat Q+\hat {\cal Y} +1)/2$
and width $\hat w = (\hat Q-\hat {\cal Y} +1)/2$ of the tableau:
\bea
\begin{array}{rcl}
n_f^* \equiv  h&=& n_f + \frac{1}{Q}\theta\theta\dg,\cr
2S \equiv w &=& n_b +\frac{1}{Q}\theta\dg \theta
\end{array}\parbox{2cm}{  
\begin{picture}(100,40)(0,0)
\multiput(20,25)(5,0){6}{\framebox(5,5){}}
\put(30,37){\vector(-1,0){10}}
\put(40,37){\vector(1,0){10}}
\put(31,35){$2S$}
\multiput(20,20)(0,-5){4}{\framebox(5,5){}}
\put(13,22){\vector(0,1){8}}
\put(10,17){$n_f^*$}
\put(13,13){\vector(0,-1){8}}
\end{picture} }
\eea
For
the fundamental representation,  described by the state
\bea
\vert \si \ra = f\dg_{\si}\vert 0 \ra \equiv 
b\dg_{\si}\vert 0 \ra
\parbox{1cm}{\begin{picture}(40,20)(0,0)
\put(20,5){\framebox(12,12)}
\put(25,22){\vector(-1,0){5}}
\put(24.5,20){$ 1$}
\put(27,22){\vector(1,0){5}}
\put(15,14){\vector(0,1){3}}
\put(13.5,9){$1$}
\put(15,8){\vector(0,-1){3}}
\end{picture}
}\label{newc}
\eea
$n_f^*=2S =1$.  Independently
of the way we represent the spin, some bosonic and fermionic
character is always present, reflecting the fact that a spin can
give rise to a ``bosonic'' local moment, or it can produce ``fermionic'' singlet bound-states. 
Traditionally,  one of the above constraints is
dropped: in the approach now adopted, both constraints are simultaneously
applied.

There are two ways in which we can use the new constraint. We can
work within a ``grand canonical'' ensemble, where $\hat Q$ is fixed, but
$\hat{\cal Y}$ is associated with a chemical potential, 
\bea 
H' = H + \zeta { \cal Y}
\eea
By tuning $\zeta$ from negative, to positive values,  the ensemble is driven from 
an  antisymmetric to a symmetric representation. 
In fact, since the cazimir ${\bf S} ^2= \hat Q ( N - \hat{\cal Y}-
\frac{\hat Q}{N})$  
is linearly related to $\hat {\cal Y}$, a finite 
value of $\zeta$ is physically equivalent to the introduction
of a 
Hund's interaction into the Hamiltonian.
\bea
H'= H - \frac{1}{Q} \zeta {\bf S} ^2,
\eea
where a constant term $\zeta(N- Q/N)$ has been omitted. 
The supersymmetric spin representation thus enables us to
progressively increase the strength of the magnetic interactions
by tuning the spin representation. 

Alternatively, we may work with a definite representation,
where $\hat{\cal Y} = Y$.
The partition function for this model
is 
\bea
Z[Q_o, Y]
= { \rm Tr}[ P_{Q_o, Y} e^{ - \beta H }]
\eea
where $P_{Q_o, Y}$ projects out the states with 
definite $\hat Q=Q_o$ and $\hat{\cal Y} = Y$. By specifying these
two constraints, we are still 
working in an ensemble where the individual
number of fermions or bosons are not separately constrained,
and in this way, we are able to develop a supersymmetric
field theory. 
We can implement
these two constraints by carrying out a Fourier transform
over the chemical potentials $\lambda$ and $\zeta$ associated 
with $\hat Q$ and $\hat {\cal Y}$ respectively,
\bea
Z[Q_o, Y]= \int \frac{d\lambda d \zeta}{(2 \pi i T)^2}
{ \rm Tr}[ e^{ - \beta (H+\lambda(\hat Q- Q_o)+ \zeta( \hat{\cal Y}
- Y) }]
\eea
where both $\zeta= \lambda_o+ ix$ and $\lambda= \lambda_o + i y$
are integrated along an imaginary axis, $x, y \in [ 0 , 2 \pi
T]$.

\section{Application to the Under-screened Kondo model}

\subsection{Formulation of Lagrangian}

To illustrate the approach, we develop it for the
single impurity Kondo model, given by
\newcommand\vp{\vphantom{\frac{J}{N}}}
\bea
H= \overbrace{\vp\sum_{k, \alpha} \epsilon_{k} c\dg_{k \alpha}c_{k \alpha
}}^{H_o} +\overbrace{\frac{J}{N} c\dg_{\alpha}\pmb{$\Gamma$}_{\alpha \beta}c_{\beta}
\cdot
{\bf S}}^{H_K}
+ \overbrace{\vp
\lambda\hat Q}^{H_Q}
+ \overbrace{\vp
\frac{\zeta}{Q_o}\hat Q \hat { \cal Y}}^{H_{Y}}
\eea
Here, $H_o$ describes the conduction electron sea, $H_K$ is
the interaction between the  conduction electron spin density,
 and the local moment, where
$c\dg_{\alpha} = {n_s}^{-1/2} \sum_{k} c\dg_{k\alpha}
$ creates an electron at the site of the local moment ($n_s$ = no. 
of sites). $H_Q$ and $H_Y$
impose the constraints.  (Note
the  way in which $H_Y$ has been written: by multiplying the operator
$ \hat{\cal Y}$ by $\hat Q$, we cast it in a form which is
unchanged upon normal ordering:
$\hat Q \hat { \cal Y}= :\hat Q { \cal
Y}:$. By 
writing it in this way we can immediately 
translate the fields to their coherent state
representation. Inside a path-integral, where $Q=Q_o$ is imposed,
we are then 
able to replace
$\frac{1}{Q_o}Q\hat{\cal Y}\rarrow { \cal Y}$.)

To date, the one-channel 
Kondo model has  been studied in the large
$N$ approach for completely symmetric\cite{georges} 
and  completely antisymmetric\cite{readimp}
representations of $\bf
S$.  For the latter
the local moment is
quenched to form a local Fermi liquid;  for
symmetric representations
the 
spin $S$ 
is only partially screened by the Kondo effect to form a spin $S-\frac{1}{2}$. 
By tuning the representation from the one limit to the other,
we are able to examine how the local Fermi liquid interacts with the emergent local moment as the local moment grows. 
For fully symmetric representations, it is known that the
residual Fermi liquid decouples from the partially
screened moment\cite{blandin}. One of the surprising
discoveries of this study, is that in
intermediate representations,  the heavy Fermi
liquid and the partially screened moment can
become antiferromagnetically  coupled
and the strong-coupling fixed point becomes
unstable. We shall see that in this simple model,
a  two-stage Kondo effect then takes place; 
richer consequences are likely in a lattice
model. 

Our first step is to write the constrained partition function as 
a path integral 
\bea
Z= \int { \cal D} [ c,f,b,\lambda, \zeta]
e^{- \int _0^{\beta}\bigl[{\cal L}_o+H_K+{\cal L}_{susy}-(\lambda Q_o+ \zeta Y)\bigr] d
\tau}
\eea 
where we have divided the action into three terms:

\bea
{ \cal L}_o &=& \sum_{\bf k \si}
c\dg_{\bf k \si} (\partial_{\tau}+ \epsilon_{\bf k})c_{\bf k \si}, \cr\cr
H_K&=& -\frac{J}{N} \sum_{ \alpha, \beta}
\biggl[
f\dg_{\alpha}{c}_{\alpha}
{c}\dg_{ \beta}f_{\beta}
+b\dg_{\alpha}{c}_{\alpha}
{c}\dg_{ \beta}b_{\beta}
\biggr], \cr\cr
\Ls &=& \sum_{\si}[f\dg_{\si} \partial_{\tau}
f_{\si} + b\dg_{\si} \partial_{\tau}
b_{\si}] + H_Q+H_Y.
\eea
The first term
describes the conduction electrons, in the second term we have
rewritten the local spin in terms of slave-fields, 
and the third term contains the machinery of the
supersymmetric representation. We use a single notation
for the field operators and the c-number fields that represent
them inside the path integral. 

Our next step is to formulate the Lagrangian in a form that
clearly exhibits the supersymmetry. We shall
begin by casting 
$\Ls$ in a form which is gauge invariant under
time-dependent super-rotations.  
It is 
convenient to combine the slave fields into a single spinor,
\bea
\Psi_{\si }= 
\left(
\matrix{f_{\si} \cr
 b_{\si}}
\right), \qquad
\Psi\dg_{\si}= \bigl(f\dg_{\si},
\  b\dg_{\si} \bigr).
\eea
Using this notation,
\bea
\Ls = \sum_{\si}\Psi\dg_{\si}\biggl[
\partial_{\tau} + \lambda + \zeta \tau_3\biggr]
\Psi_{\si} - \frac{2 \zeta}{Q_o} \theta\dg\theta 
\eea
where $\tau_3$ is a Pauli matrix. 
Since the starting Hamiltonian and each of the constraints
commutes with the super-generators,
the full Lagrangian is invariant (appendix B) under time-independent
super-rotations 
$
\Ps \rarrow g \Ps
$
where
\bea
g = \mat{\sqrt{1 - \eta \bar \eta} & \eta\cr -\bar \eta & \sqrt{1- \bar \eta
\eta}}\label{superrot}
\eea
is an element of the supergroup $SU(1\vert 1)$. 
If we make this transformation time dependent, the derivative
terms become 
\bea
\Ps \dg \D \Ps \rarrow  \Ps \dg [ \D + ( g \dg \D g)]\Ps.
\eea
Expanding the second term, we obtain
\bea
\Ps \dg (g \dg \D  g)  \Ps = \theta \dg \D \eta + \bar \eta \D \theta
+ {Q_o}\bar \eta \D \eta
\label{intx}
\eea
where we have replaced $\Ps \dg \Ps \rarrow Q_o$ inside
the path integral. 
Since $Z$ is unchanged by this change of basis, we
can integrate over
all $g(\tau)$ 
\bea Z&=& \int D[\bar \eta, \eta] 
\int D [ c,f,b,\lambda, \zeta]
e^{- \int _0^{\beta}\bigl[{\cal L}+ \theta \dg \D \eta + \bar \eta \D \theta
+ {Q_o}\bar \eta \D \eta \bigr] d
\tau}\cr
&=&
\int D [ c,f,b,\lambda, \zeta]
\exp\biggl\{- \int _0^{\beta}\bigl[{\cal L}-\frac{1}{Q_o} 
\theta\dg\D  \theta \bigr]
d
\tau\biggr\}
\eea 
By absorbing the additional term into a redefined 
\bea
\Ls^* = \sum_{\si}\Psi\dg_{\si}\biggl[
\partial_{\tau} + \lambda + \zeta \tau_3\biggr]
\Psi_{\si} - \frac{1}{Q_o} 
\theta\dg(\D + 2 \zeta) \theta 
\eea
the Lagrangian becomes invariant under time-dependent super-rotations. 
The first term in $\Ls^*$ describes the level
splitting between the bosonic and fermionic components
of the spin. The second term describes a residual interaction
between the spin and heavy electron fluid. 
We can factorize this term, to obtain
\newcommand\Ds{(\D + 2 \zeta)}
\newcommand\Fb{f\dg_{\si } b_{\si}}
\newcommand\bF{b\dg_{\si } f_{\si}}
\bea
- \frac{1}{Q_o} 
\theta\dg(\D + 2 \zeta) \theta \rarrow
Q_o\alpha \dg \Ds \alpha+  
\overbrace{\bigl[
\Fb \Ds \alpha + \alpha \dg\Ds \bF\bigr]}
^{H_I} 
\eea
The first term tells us that
the field $\alpha$  represents a { \sl dynamical}
fermion with the commutation
algebra $\{\alpha , \alpha\dg\} = 1/Q_o$ . This spin-less particle 
mediates the interaction between the spin 
and the  Fermi liquid; $H_I$ 
defines the vertex for the decay process
$
f^-_{\si}\rightleftharpoons b_{\si} + \alpha^-.
$
To 
represent this process, we denote the propagator for the $\alpha$ fermion
by the Feynman diagram 
\bea
\raisebox{-0.2cm}[0cm][0cm]{\epsfig{file=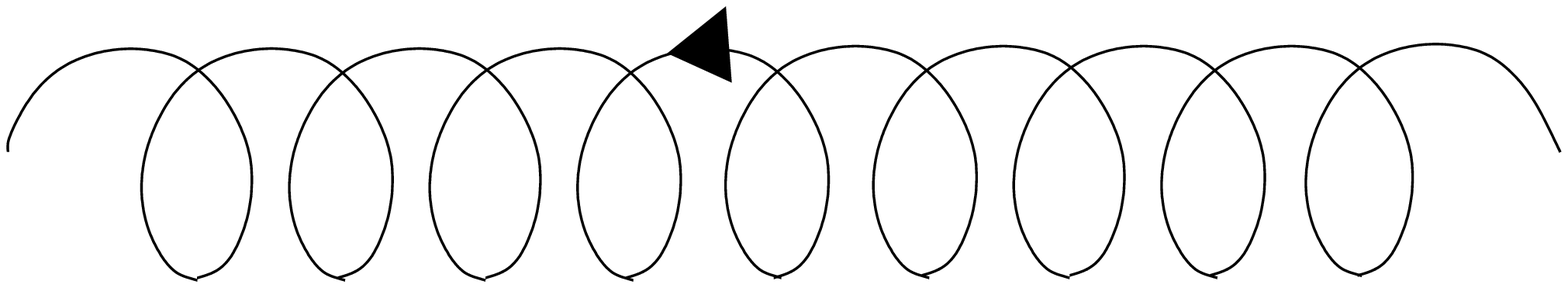,width=2.5cm}} &=& [Q_o(i \omega_n - 2 \zeta)]^{-1}
\eea
The vertices which inter-convert the heavy electron and  spin bosons
will be denoted by 
\bxwidth=2.5truecm
\bea
\left.
\begin{array}{r}
\raisebox{-1cm}[0cm][1.2cm]{\frm{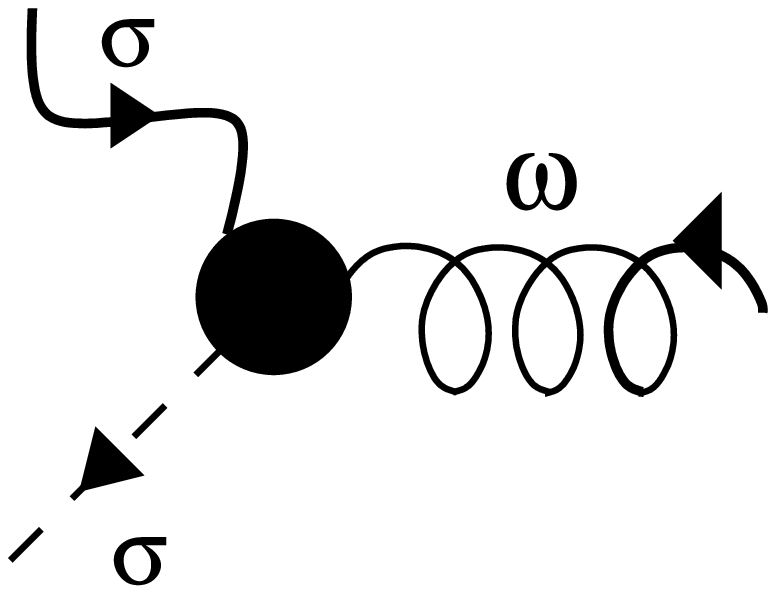}}\cr
\raisebox{-1cm}[0cm][1.2cm]{\frm{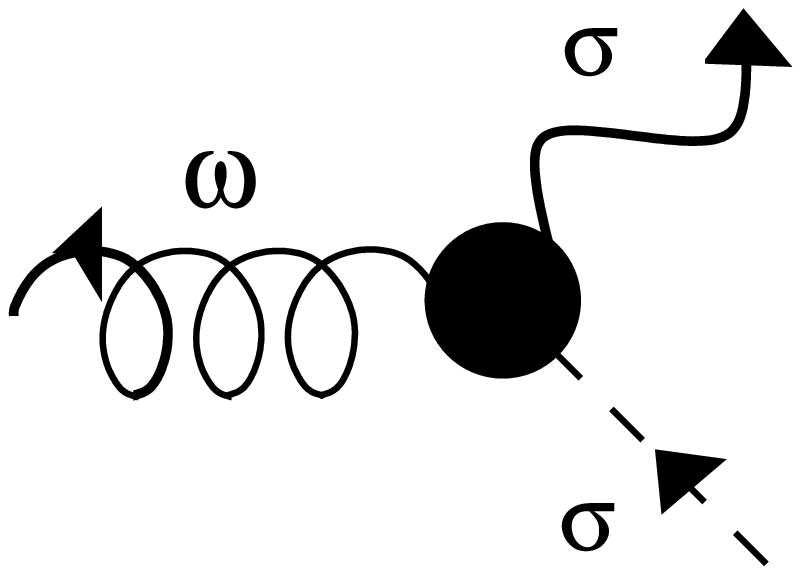}}
\end{array}\right\}
=
i(\D + 2 \zeta)\equiv i(2 \zeta - i \omega_n),
\eea
where the ``$i$'' is required to give the correct amplitude for
the exchange of the gauge fermion and 
\bea
\frm{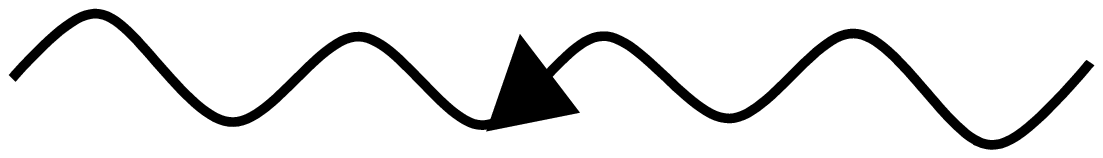}&=&G_b(i \nu_n) = ( i \nu_n - \lambda_b)^{-1}\cr
\frm{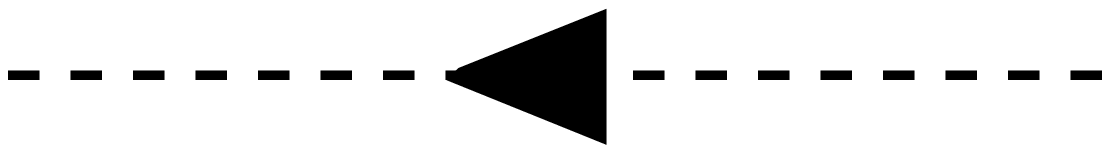}&=&G_f(i \omega_n) = ( i \omega_n - \lambda_f )^{-1}
\eea
represent the propagators for spin bosons and the f-electrons.
The mediated bare interaction
between the spins and the heavy f-electrons  is then
\bea
\raisebox{-0.6cm}[0cm][0.9cm]{\epsfig{file=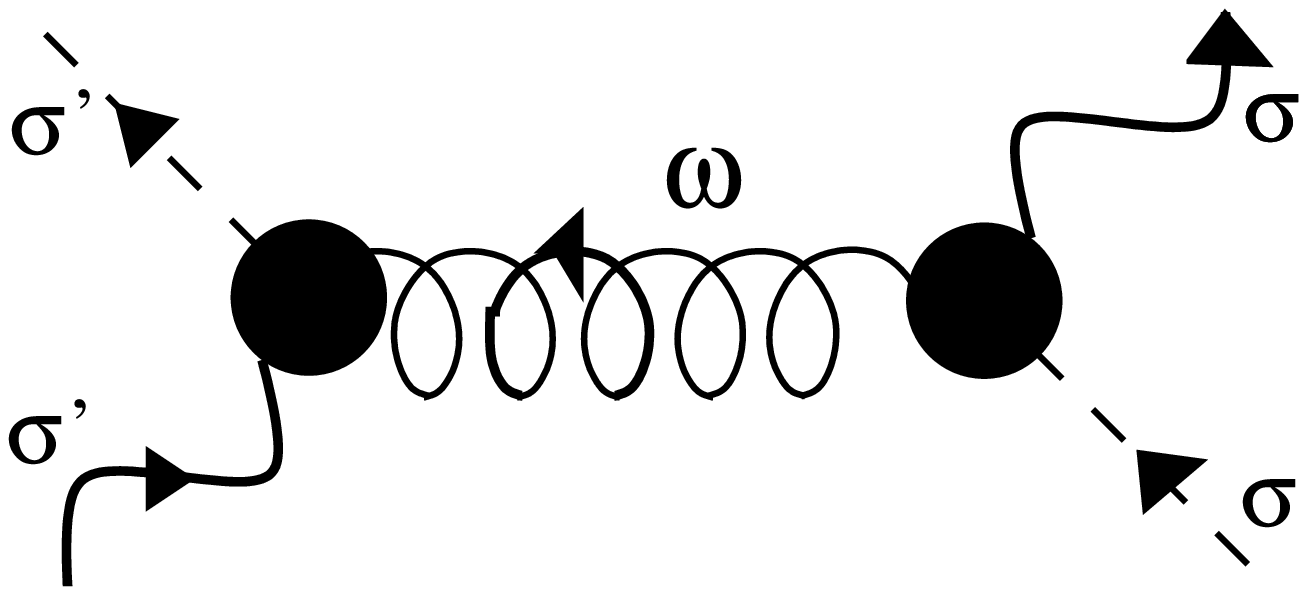,width=3.5cm}}= \frac{1}{Q_o}(2 \zeta-\omega)
\eea 
It is a rather unique feature of this kind of approach
that the spin interactions are carried by fermions, rather than
bosons. 
Our final form for $\Ls^*$ can now be compactly written 
\newcommand\slp{{\raise.15ex\hbox{$/$}\kern-.57em\hbox{$\partial$}}}
\bea
\Ls^* =  \sum_{\si}\Psi\dg_{\si}\biggl(
\partial_{\tau} +\lambda+\mat{\zeta &
\slp \alpha\cr 
\bar \slp\alpha \dg &
-\zeta}
 \biggr)
\Psi_{\si} +
Q_o\alpha \dg\slp \alpha 
\eea
where  we have defined
$(\D + 2 \zeta)\equiv \slp$, 
$(-\D + 2 \zeta)\equiv \bar\slp$. 

As our next step, we factorize the Kondo interaction term $H_K$
\newcommand \cV{ {\cal V}}
\bea H_K \rarrow H_K^*=\sum_{\si}
 \biggl[ \bigl(c\dg_{\si}\bar V f_{\si}+ V f\dg_{\si}c_{\si}\bigr)
+ \bigl( c\dg_{\si} \bar \phi b_{\si} + 
\phi b\dg_{\si}c_{\si}\bigr)  \biggr] + \frac{N}{J}( \bar V V + \bar \phi \phi)
\eea
where $V$ is a complex c-number field and $\phi$ is its
fermionic partner. If we now introduce 
\bea
\cV = \mat{V\cr \phi}, \ \cV \dg = ( \bar V, \bar \phi)
\eea
then the transformed Lagrangian takes the form
$
{\cal L} =  { \cal L}_o + \Ls ^* + H_K^*
$,
where 
\bea
{ \cal L}_o &=& \sum_{\bf k \si}
c\dg_{\bf k \si} (\partial_{\tau}+ \epsilon_{\bf k})c_{\bf k \si}, \cr\cr
H_K &=& \sum_{\si }\bigl[
\Ps \dg \cV c_{\si} + c\dg_{\si} \cV \dg \Ps\bigr] + \frac{N}{J}\cV \dg \cV\cr
\cr
\Ls^* &=&  \sum_{\si}\Psi\dg_{\si}\biggl(
\partial_{\tau} +\lambda+\mat{\zeta &
\slp \alpha\cr 
\bar \slp\alpha \dg &
-\zeta}
 \biggr)
\Psi_{\si} +
Q_o\alpha \dg\slp \alpha 
\eea
Let us briefly examine the gauge invariance of this Lagrangian.
If
\bea
h
= ge^{i (\theta_Q+\theta_{\zeta}\tau_3)},\eea
is a general $SU(1|1)$ matrix, where $g$ takes the form (\ref{superrot}), 
then under the gauge transformation
$
\Ps\rarrow h \Ps
, {\cal V} \rarrow h {\cal V},
$,
${\cal L}_o$ and $H_K$ are invariant, but $\Ls^*$ becomes
\bea
\Ls^* \rarrow \sum_{\si}\Psi\dg_{\si}h\dg\biggl(
\partial_{\tau} +\lambda+\mat{\zeta &
\slp \alpha\cr 
\bar \slp\alpha \dg &
-\zeta}
 \biggr)
h\Psi_{\si} +
Q_o\alpha \dg\slp \alpha 
\eea
When we expand the first term (Appendix B), we find that
\bea
\Ls^*[\lambda, \zeta, \alpha, \alpha\dg]
\rarrow
\Ls^*[\lambda', \zeta', \alpha', \alpha^{\dagger '}]
\eea
where $\lambda' = \lambda + i \dot \theta_Q$, $\zeta'=\zeta + i \dot
\theta_{\zeta}$ and $\alpha' = (\alpha +
\eta)e^{-2 i \theta_{\zeta}}$,  so $\cal L$ is
invariant under the gauge transformation, 
\bea
\Ps&\rarrow &h \Ps
, {\cal V} \rarrow h {\cal V},\cr
\lambda &\rarrow &\lambda - i \dot \theta_Q, \ 
\zeta \rarrow \zeta - i \dot \theta_{\zeta},\cr
\alpha &\rarrow &e^{2 i \theta_{\zeta}}\alpha -
\eta, \ 
\alpha\dg \rarrow e^{-2 i
\theta_{\zeta}}\alpha\dg - \bar \eta.  
\label{fullg}
\eea

This gauge invariance leads to bosonic and fermionic
zero-modes. To eliminate them, we must 
carry out a gauge fixing procedure. We can always parameterize
${\cal V}$ in the form
\bea
\cV = h\mat{V_o\cr 0}= g\mat{V_o
e^{i(\theta_Q+ \theta_{\zeta})}\cr 0} , 
\eea
or written out explicitly,
\bea
\left(
\matrix{V\cr \phi
}
\right)= V_o e^{i \theta_f}\left(
\matrix{ \sqrt{1 - \eta \bar \eta} \cr - \bar \eta
}
\right).
\eea
where $\theta_f= \theta_Q+ \theta_{\zeta}$ and $V_o= \sqrt{\bar V
V +
\bar
\phi
\phi}$ is real. This transformation uniquely specifies both
$
\bar \eta = - \frac{V_o}{V} \phi
$
and the phase factor and $e^{i \theta_f}= \frac{V}{\vert V\vert
}$, but  does not specify $\theta_b= \theta_Q-
\theta_{\zeta}$. We shall adopt a gauge choice where $\theta_b=
0$.   By
applying the gauge transformation (\ref{fullg}), we can  absorb
the fermionic fluctuations in $\cal V$ into a redefinition of the
fields.  The gauge fixed hybridization is now
\bea
H^*_K= V_o\sum_{\bf \vec \si}
 \biggl[ c\dg_{\si}f_{\si}+  f\dg_{\si}c_{\si}\biggr]
 + \frac{N V_o^2}{J}.
\eea
With our gauge choice $\theta_b=0$, the variable
$\lambda_f(\tau) = \lambda+ \zeta + i \dot \theta_f$ becomes
 {\sl dynamical} , but the variable $\lambda_b= \lambda- \zeta
 + i \dot \theta_b= \lambda- \zeta$ remains a time-independent
integration variable. 
In this gauge fixed form of the Hamiltonian, the
interaction between the Fermi and spin fluids is entirely
contained within
$\Ls ^*$, and it is here where we should look if we are to obtain
new physics. 

In the existing large N approach to the Kondo model, magnetic
interactions are a $1/N^2$  correction to the mean-field theory, an
order of accuracy that is beyond current theoretical approaches. 
A supersymmetric approach enhances the magnetic interactions
by a factor of $N^2$, 
bringing them within the realm of 
Gaussian fluctuations about a new large $N$
mean-field theory.  To carry out concrete calculations, we expand around
a large $N$ saddle point of the path integral obtained by taking $N\rarrow
\infty$, maintaining
$Q/N= q$ and $Y/N $ fixed. By allowing the number of bosons $n_b = N
\tilde n_b$ to become  large, the Bose field is able to condense and form a
magnetic moment:
\bea
\la b_{\si}\ra =\sqrt{M} z_{\si}\sim O(\sqrt{N})
\eea
where $z_{\si}$ is a unit spinor. The magnetic interaction between
spins at different sites is given by RKKY diagrams (Fig. \ref{figex}).  The factors of $\sqrt{N}$ associated with the Bose
condensate produce an the inter-site magnetic interaction of order
$O(1)$ : a factor of $N^2$ enhancement. 
These magnetic  corrections appear as part of the 
Gaussian fluctuations of the $\alpha$ fermion 
and by calculating them, we are 
able to carry out a controlled  treatment of magnetism and the
Kondo effect. 
\fight=\textwidth
\fight=0.6\textwidth
\setcounter{figure}{1}
\begin{figure}[tb]
\epsfxsize= 0.6\textwidth
% ***********For one column  ********************
\centerline{\epsfbox{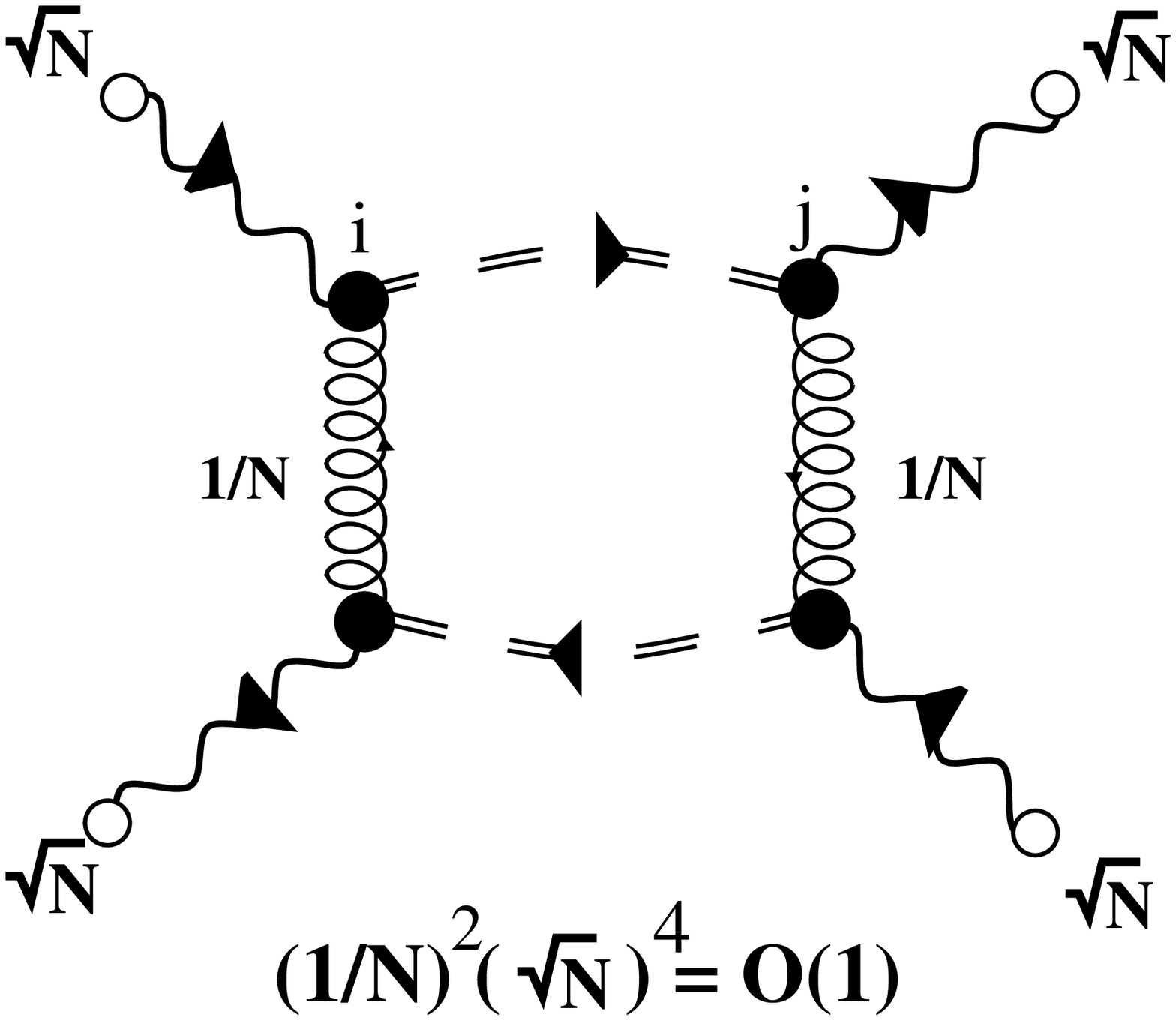}}
%\epsfxsize=7.0in 
% ***********************************8
\vskip 0.2truein
\protect\caption{Magnetic 
interaction between two spins at site $i$ and $j$ within
the current approach.  
(Note that when $V_o>0$, the f-fermion
can propagate from site to site.) 
The factor
$(\sqrt{N})^{4}$ associated with the Bose condensate, 
enhances the magnetic interaction by a factor of $N^2$, 
so that it appears as the 
as the first in a  series
of RPA diagrams associated with the Gaussian
fluctuations of the $\alpha$ fermion. 
}
\label{figex}
\end{figure}

\subsection{Mean Field Theory}

To illustrate this kind of calculation, we  develop the machinery
for the single impurity model. Although there are no inter-site
magnetic interactions, the machinery of the supersymmetry is
needed to compute the magnetic
interaction between the partially screened local moment
and the Fermi liquid in the 
single impurity model.  The techniques that we now illustrate can
be generalized to the lattice. 

Our first step is to formally integrate out the conduction($c$) and the slave
fields ($f,b$) 
\bea
Z&=& \int D[\alpha, V, \lambda, \zeta]e^{- S_{eff}}
\cr
e^{- S_{eff}}&=&  \int D[c,f,b]e^{- \int_0^{\beta}({\cal L}_o+
{\cal L}^*_{susy}+H_K) d \tau}
\eea
Since the second integral is bilinear in the fields, it can be carried
out to yield
\bea
S_{eff}[\alpha,\lambda,\zeta] = - {\rm STr } {\rm ln} \bigl[ \partial_{\tau}
+ \lambda  + \zeta \tau_3 + \underline{\Sigma}\bigr]+  \int_0^{\beta}\bigl[ Q_o \alpha \dg \slp \alpha - \lambda Q_0 - \zeta
Y\bigr]d \tau
\eea
and
$
\underline{\Sigma}$
is the self-energy correction induced by the coupling to
the conduction electrons and the $\alpha$ fermion and
$STr[ \dots]$ denotes the ``super trace''- the difference
between the fermionic and the bosonic trace:
\bea
{\rm STr} \left[
\matrix{ F & \alpha \cr \beta & B
}
\right]= {\rm Tr}[F] - {\rm Tr} [B]
\eea
Our procedure is then to
expand the effective action to quadratic order around 
the saddle point where
$\lambda$, $\zeta$ and $V(\tau)= V_o $ are static, and
$\alpha=0$. 
\bea
S_{eff}[V, \lambda,\zeta, \alpha]= S_{MF}
+ O(\delta \Lambda^2)
\eea
where
$\delta \Lambda$ denotes the fluctuations. $S_{MF}=\beta F_{MF}$
determines the leading $O(N)$
mean-field contribution to the Free energy. By 
carrying out the Gaussian integral over the fluctuations in 
$\delta \Lambda$ we can then determine the $O(1)$ correction to the
mean-field theory and 
the magnetic interactions within the medium.

We begin by computing the saddle point, 
described by the 
the mean-field Hamiltonian
\bea
H_{mft}= \sum_{\bf k \si}
\epsilon_{\bf k}c\dg_{\bf k \si}c_{\bf k \si}+
V_o\sum_{\si}[c\dg_{\si}f_{\si}+ f\dg_{\si} c_{\si}]
+ ( \lambda - \zeta) n_b + ( \lambda + \zeta) n_f -\lambda Q_o
\eea
where $V_o$ and $\lambda$ are to be determined self-consistently.
This mean-field theory describes a Kondo resonance formed between
the conduction electrons and the antisymmetric part of the
spin.  The residual symmetric part of the spin is unquenched,
and described by a sharp bosonic level at energy $\lambda_b = \lambda- \zeta$.
For the moment, we shall work in the ensemble of definite $\zeta$,
examining how the mean-field evolves as we increase $\zeta$ to favor
representations that are increasingly symmetric. 
If we define $\lambda_f = \lambda + \zeta$, $\lambda_b =\lambda - \zeta$,
then in the presence of the finite hybridization, 
the Greens functions for the slave fields are now given by
\bxwidth=2.5truecm
\bea
\frm{bprop.ps}&=&G_b(i \nu_n) = ( i \nu_n - \lambda_b)^{-1}\cr
\frm{fprop.ps}&=&G_f(i \omega_n) = ( i \omega_n - \lambda_f + i \Delta {\rm sign}(\omega_n))^{-1}
\eea
where $\Delta = \pi V_o^2 \rho$ is the hybridization width of the conduction
electron and $\rho$ is the conduction electron density of states. 
The mean-field Free energy is then given by
\bea
F_{mft}= NT \sum_n \biggl(\ln [ - G^{-1}_b( i \nu _n)] - \ln [ - G^{-1}_f( i \omega _n)] \biggr) e^{i \omega_n 0^+} + N \frac{V_o^2}{J}- \lambda N q
\eea
The first term is just the free energy of a free boson. 
Carrying out the frequency sum on the second term, (Appendix C) we
obtain
\bea
\frac{F_{mft}}{N}= \Phi_f(\xi) + \Phi_b(\lambda_b)
+\frac{V_o^2}{J} - \lambda q
\eea
where
\bea
\Phi_f(z) &=& - 2 T {\rm Re}\ln  \left[ \frac{
\Gamma\bigl( \frac{z+D}{2 \pi i T} \bigr)}{\Gamma\bigl( 
\frac{z}{2 \pi i T}+ \frac{1}{2}\bigr)}
\right]-D/2\cr
\Phi_b(z) &=&  T\ln(1- e^{ - \beta z})
\eea
are the fermionic and bosonic contributions to the energy, 
$D$ 
is the conduction band half-width and $\xi = \lambda_f + i \Delta$. 
If we differentiate this
result with respect to $\Delta$ and $\lambda$, we obtain
\bea
\frac{\pi}{N}\left(\frac{\partial F}{\partial \Delta}+ i\frac{\partial F}{\partial \lambda }
\right)&=&
\psi\biggl(\frac{\xi}{2 \pi i T}+ \frac{1}{2}
 \biggr)- \ln \biggl( 
\frac{D}{2 \pi i T}\biggr)+ \frac{1}{J \rho}+i \pi(q-\tilde n_b)\cr
&=& \psi\biggl(\frac{\xi}{2 \pi i T}+ \frac{1}{2}
 \biggr)-\ln \biggl( 
\frac{T_Ke^{i \pi ( q- \tilde n_b)}}{2 \pi i T}\biggr)=  0
\eea
where $\psi(z)= \partial_z \ln \Gamma(z)$ is the digamma function, 
$\tilde n_b= n_b/N=[ \exp( \lambda_b / T)-1]^{-1}$ and 
$T_K= D \exp[ - 1/ J \rho_o]$ is the Kondo temperature.
At zero temperature we may replace, 
$\psi(z)\rarrow \ln z$, so the $T=0$ 
mean-field equations are then
\bea
\xi = T_K e^{i \pi ( q- \tilde n_b)}
\eea
\begin{figure}[tb]
\epsfxsize= 0.6 \textwidth
% ***********For one column  ********************
%\epsfxsize=7.0in 
% ***********************************8
\centerline{\epsfbox{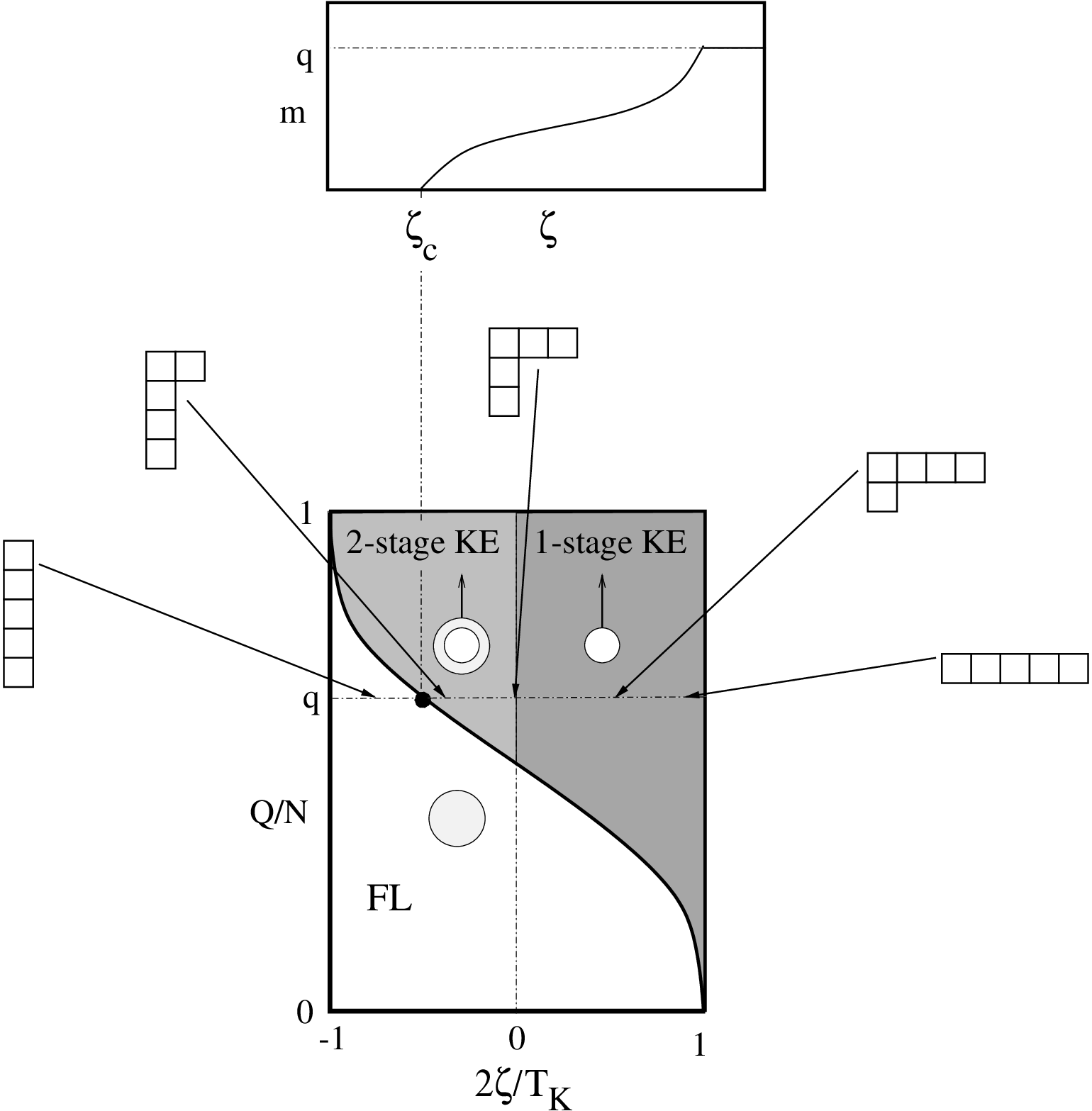}}
\vskip 0.1truein
\protect\caption{
Top: magnetization
of partially screened local moment as a function of $\zeta$. 
Bottom: Phase diagram for the supersymmetric impurity Kondo
model, showing how the representation of the local moment evolves
as $\zeta$ is increased. Shaded area indicates under-screened region.  For $\zeta<0$, the
under-screened region involves  a two-stage Kondo effect. 
}
\label{fig1}
\end{figure}
If the Bose field condenses, then $\lambda_b= 0$,
so $\lambda= \zeta$, $\lambda_f=2 \zeta$ and 
$\xi = 2 \zeta + i \Delta$. In this case, we can solve for
the size of the unquenched moment, $M= N m = N\tilde  n_b$ and 
the width $\Delta$ of the  Abrikosov Suhl resonance with which it
co-exists:
\bea
m & =& {q} - \frac{1}{\pi} \cos^{-1} \left( 
\frac{2 \zeta}{T_K}\right),\qquad (\zeta> \zeta_c)\cr
\Delta &=& \sqrt{(T_K)^2 - (2 \zeta)^2},
\eea
where 
\bea
\zeta_c= (T_K/2)\cos (\pi q).\eea 
corresponds to a critical
value of the Hund's interaction beyond which the local moment
develops an unscreened component. 
There are three regions:
\bea
m = \left\{
\begin{array}{lc}
{q} & 2 \zeta/T_K > 1\cr
{q} - \frac{1}{\pi} \cos^{-1} \left( 
\frac{2 \zeta}{T_K}\right),&1>2\zeta/T_K > \cos(\pi q)\cr
0 &\cos(\pi q)> 2\zeta/T_K
\end{array}
\right.
\eea
corresponding to an unscreened, partially screened and fully screened
local moment.  
Fig. 1.
shows the mean-field phase diagram. 

Next, let us consider the residual interactions between the unquenched
local moment, and the Fermi liquid.  The constraint term 
$ \zeta\cal Y$ generates the residual interaction
\bea
H_I= - \frac{2 \zeta}{Q_o} 
( f\dg_{\si} b_{\si}) ( b\dg_{\si'}f_{\si'})
\eea
between the heavy-f electron and the unscreened moment. 
Conventional wisdom, based on the spin $S$ Kondo
model supposes that such residual interactions are always Ferromagnetic.
For $\zeta >0$, this is indeed the case,
and the fixed point described by the
mean-field theory is thus stable.  By contrast, if $\zeta<0$, then
the residual interaction is {\sl antiferromagnetic}. The presence of
such terms is unexpected. We shall see that for the impurity,
this leads to  a two stage Kondo effect.
In the single impurity model, the condition $\zeta<0$ corresponds to
the requirement that $n_f^*> N/2$, in other-words, the requirement that
the antisymmetric component of the spin representation is more than
half-filled. 

\subsection{Calculation of the Magnetization using Gaussian
Fluctuations}

One way to examine the consequences of this 
residual coupling on the single-ion  Kondo effect is to compute the
field-dependent magnetization of the ground-state. The application of
a field provides a controlled way of examining the cross-overs
associated with screening processes.
To compute the magnetization, we need to introduce a magnetic field
and calculate the field dependent ground-state energy, including the
effect of the Gaussian fluctuations around the mean-field theory.
For $SU(N)$, there are $N-1$ ways of introducing  the magnetic
field. We shall use the form 
\bea
H_B = -B \sum_{\si} m_{\si}(n_{f\si}+n_{b\si})
\eea
where $m_1= -1$, $m_2=1$. 
With this choice, the  field splits off two  bosonic and two fermionic
states from  the other
$(N-2)$ levels.

The mean-field free
energy in a field is then given by
\bea
F_{mft}[B]=  \sum_{\si} \biggl[\Phi_f(\xi_{\si}) + \Phi_b(\lambda_{b\si}
)\biggr]
+\frac{NV_o^2}{J} - \lambda Q_o - \zeta Y
\eea
where $\xi_{\si} = \xi - B m_{\si}$, $\lambda = \lambda - B m_{\si}$
To calculate the magnetization, we must differentiate the Free energy
with respect to $B$. ( Since the Free energy  is stationary with respect
to changes in $\lambda$ and $\zeta$, we do not have to worry about how
these fields change w.r.t. $B$).
The magnetization is then 
\bea 
M &=&n_{b1} + (n_{f1}
- n_{f2})\cr &=& 2S +
\overbrace{\frac{1}{\pi}\left[ 
\tan^{-1}\left( \frac{\Delta}{-B - \lambda_f}\right)-
\tan^{-1}\left( \frac{\Delta}{B-\lambda_f}
\right)\right]}^{m_f(B)}
\eea
The first term is the  residual unquenched moment, the second term
represents the spin-polarization of the Kondo singlet. Technically, 
we should lump this term with the other $O(1)$ corrections that
we need to calculate from the  Gaussian
fluctuations. 
The mean-field only reliable predicts the terms of order $N$, and
thus to this order, 
\bea
M= 2S + O(1)
\eea
To calculate the $O(1)$ term, we need to include the zero-point corrections
to the ground-state energy. 

There are two types of Gaussian fluctuation around mean-field theory:
bosonic fluctuations in $V$, $\lambda$ and $\zeta$, plus
the fluctuations of the $\alpha$ field.  Fluctuations in $V$ and $\lambda_f=
\lambda + \zeta$
are associated with the interactions in the Fermi liquid. These
terms renormalize $m_f(B)$ and  produce an order $O(1/N)$ correction to
the magnetization.  Fluctuations in
$\lambda_b= \lambda- \zeta$ renormalize the entropy of the free moment,
and do not produce any correction to the unscreened moment. The only
$O(1)$ corrections to the magnetization are those associated with the
fermionic fluctuations. 
The Lagrangian for these fluctuations is given by
\bea
{\cal L}_{\alpha}= 
Q_o\alpha \dg \Ds \alpha+  
\overbrace{\bigl[
\Fb \Ds \alpha + \alpha \dg\Ds \bF\bigr]}
^{H_I} 
\eea
In a field the number one boson condenses, so that
$\lambda_{b1} =0$. To fulfill the
constraint on the Fermi fields,
$n_{f}^*=constant$, we require that $\lambda_f=
\lambda_{b1} + 2\zeta+B$ is field independent,
which implies that in a field, $2 \zeta= 2
\zeta_o -B$. 

When we expand the effective action $S_{eff}$ to Gaussian order
in the $\alpha field$, we obtain the correction
\bea
S_{\alpha} = -\int d1 d2 \alpha(1)D^{-1}(1-2)\alpha\dg(2)
\eea
where
\bea
-D^{-1}(1-2)&=&  Q_o( \partial_{\tau}+ 2 \zeta) -
( \partial_{\tau}+ 2 \zeta)^2\la T \theta(1) \theta\dg(2)\ra\cr
&=&  Q_o( \partial_{\tau}+ 2 \zeta) -
( \partial_{\tau}+ 2 \zeta)^2\sum G_{f\si}(1-2)G_{b\si}(2-1)
\eea
is the inverse propagator for the $\alpha$ fermion. Written out both
diagrammatically, and in the frequency domain, this is
\bea
\begin{array}{rcccl}
\biggl[\raisebox{-0.2cm}[0cm][0cm]{\frm{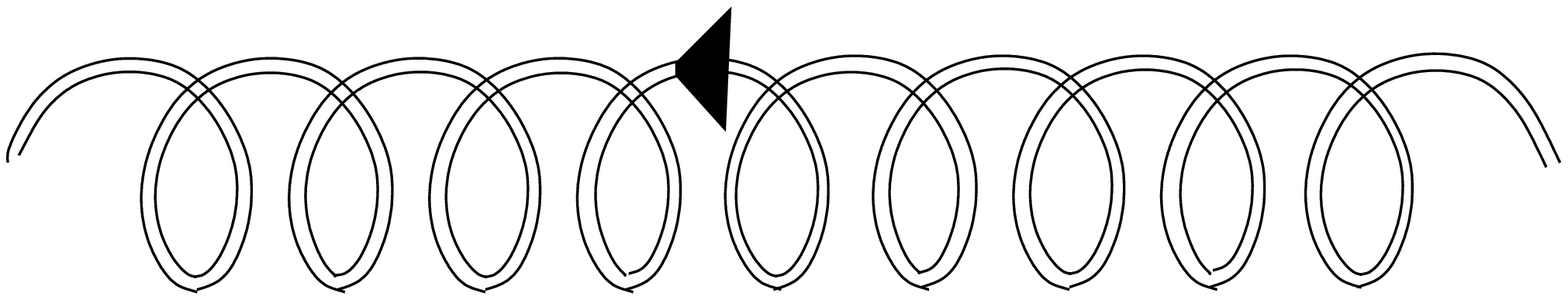}}\biggr]^{-1} &=&
\biggl[\raisebox{-0.2cm}[0cm][0cm]{\frm{gluon1.ps}}\biggr]^{-1} &-& 
\raisebox{-0.5cm}[0cm][0cm]
{\frm{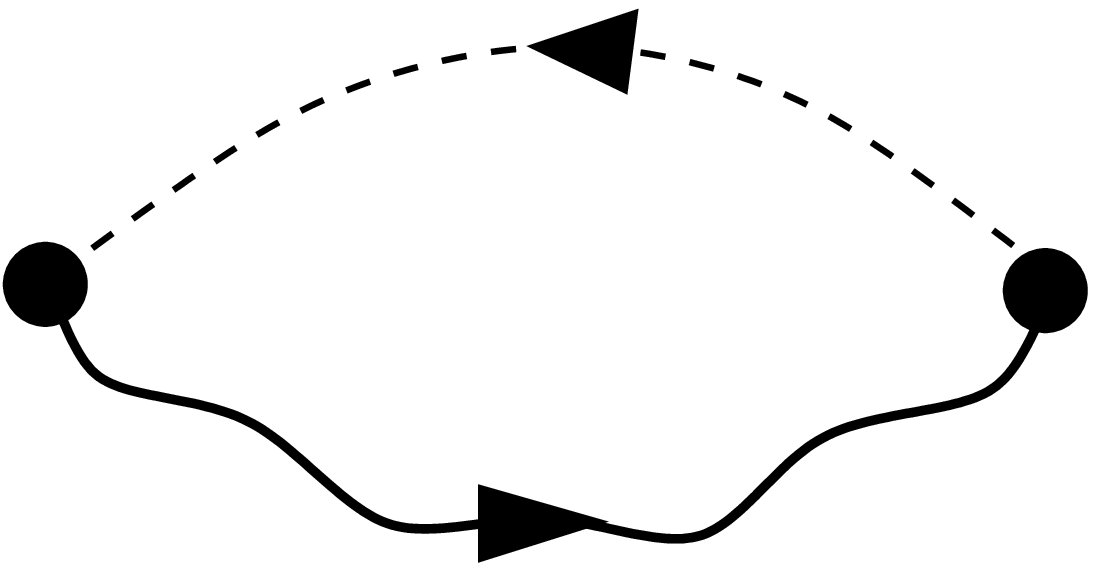}}\cr\cr
D^{-1}(\omega)&=& Q_o( \omega - 2 \zeta) 
&-& N[i(2 \zeta-\omega)^2] \Phi(\omega)\end{array}\eea
where
\bea
\Phi(\omega)= \frac{T}{N}
 \sum_{ \si \nu}
G_{f \si}( \omega+ \nu)G_{b\si} ( \nu)
\eea
It proves convenient to factorize $D^{-1}(\omega)$ as follows
\bea
D^{-1}( \omega) = Q_o ( \omega - 2 \zeta) P(\omega)
\eea
where
\bea P( \omega) = 1+ \frac{ \omega - 2 
\zeta}{q}\Phi(\omega)
\eea
A detailed calculation of $P( \omega)$ at zero
temperature ( Appendix D)
yields 
\bea
P( \omega- i \delta) = P_o(\omega+B- i \delta)
\eea
where
\bea
P_o(\omega)= 
 A_f( \omega) \left[\Delta
- \frac{ \omega - 2 \zeta_o}{\pi q} 
\left(\ln { T_K \over \om
}+ i \pi \tilde n_b\right)
\right]
\eea
is the zero-field expression for $P(\omega)$ and  
$
A_f( \omega) = Im G_f(\omega- i 
\delta)
$
is the zero-field f-spectral function.

Now the free energy  associated with the Gaussian fluctuations in $\alpha$
is given by
\bea
 F_{\alpha}&=& - T \sum_{n} \left\{
\ln\bigl[ - D^{-1}( i \om_n) \bigr] - 
\ln\bigl[Q_o( 2 \zeta- i \om_n)\bigr]
\right\}e^{i \omega_n 0^+}\cr
 &=& - T \sum_{n} 
\ln\bigl[ P(i \omega_n)e^{i \omega_n 0^+}
\bigr]
\eea
where  we subtract the Free
energy of the auxiliary $\alpha$ field in the
absence of interactions, to avoid over-counting.
Carrying this sum out by contour integration, and
taking the zero-temperature limit, the zero-point
energy is
\bea
E_{\alpha} = -\int_{-D}^{0} \frac{d \omega}{\pi}
 {\rm Im} \ln \bigl[
P_o(\omega+B-i \delta)\bigr]
\eea
where we have inserted the field dependence by
replacing 
$P(\omega)\rarrow P_o(\omega+B)$.
Differentiating this with respect to the applied magnetic field, 
the screening contribution to the magnetization due to the interaction
between the spin and Fermi fluid is given by
\bea
M_{\alpha}&=& - \frac{\partial
E_{\alpha}}{\partial B}\cr &=&\int_{-D}^0 \frac{d
\omega}{\pi} 
\partial_{\omega}{\rm Im} \ln \bigl[
P_o(\omega+B-i \delta)\bigr]
\cr
&=& \left[
 \frac{1}{\pi}{\rm
Im}
\ln
\bigl[ P_o(\omega+B-i \delta)\bigr]\right]^0_{-D}
\eea
The lower limit of this sum gives a contribution
$-\frac{1}{\pi} {\rm
Im}
\ln [ -D \ln (D/T_K) +i \pi(1+\tilde n_b)] = -1
$.
The final result for the fluctuation
contribution to
the magnetization is then
\bea
M_{\alpha}(B)
=-1 +\frac{1}{\pi} {\rm Im}\ln
\left[\Delta
+ \frac{   2 \zeta_o-B}{\pi q} 
\ln \left({ T_K e^{i \pi \tilde n_b} \over
B}\right)
\right]\eea  
To obtain the total magnetization, we must
add this to the result $M_{mft}= 2S + m_f(B)$
obtained from the mean-field theory. 
The total
magnetization, evaluated to order
$O(1)$ is then
\bea
M(B) = \overbrace{\vphantom{\left[\Delta
+ \frac{   2 \zeta_o-B}{\pi q} 
\ln \left({ T_K e^{i \pi \tilde n_b} \over
B}\right)
\right]}\bigl[2S-1 +
m_f(B) \bigr]}^{M_1(B)}+
\overbrace{\frac{1}{\pi} {\rm Im}\ln
\left[\Delta
+ \frac{   2 \zeta_o-B}{\pi q} 
\ln \left({ T_K e^{i \pi \tilde n_b} \over
B}\right)
\right]}^{M_2(B)}+O(\frac{1}{N})\eea
where 
\bea
m_f(B) = \frac{1}{\pi} \left\{\arctan\left[
\frac{2 \zeta_o +B}{\Delta}\right]-
\arctan\left[
\frac{2 \zeta_o -B}{\Delta}\right\}
\right].
\eea

{\begin{figure}[tb]\epsfysize=\fight 
\centerline{\epsfbox{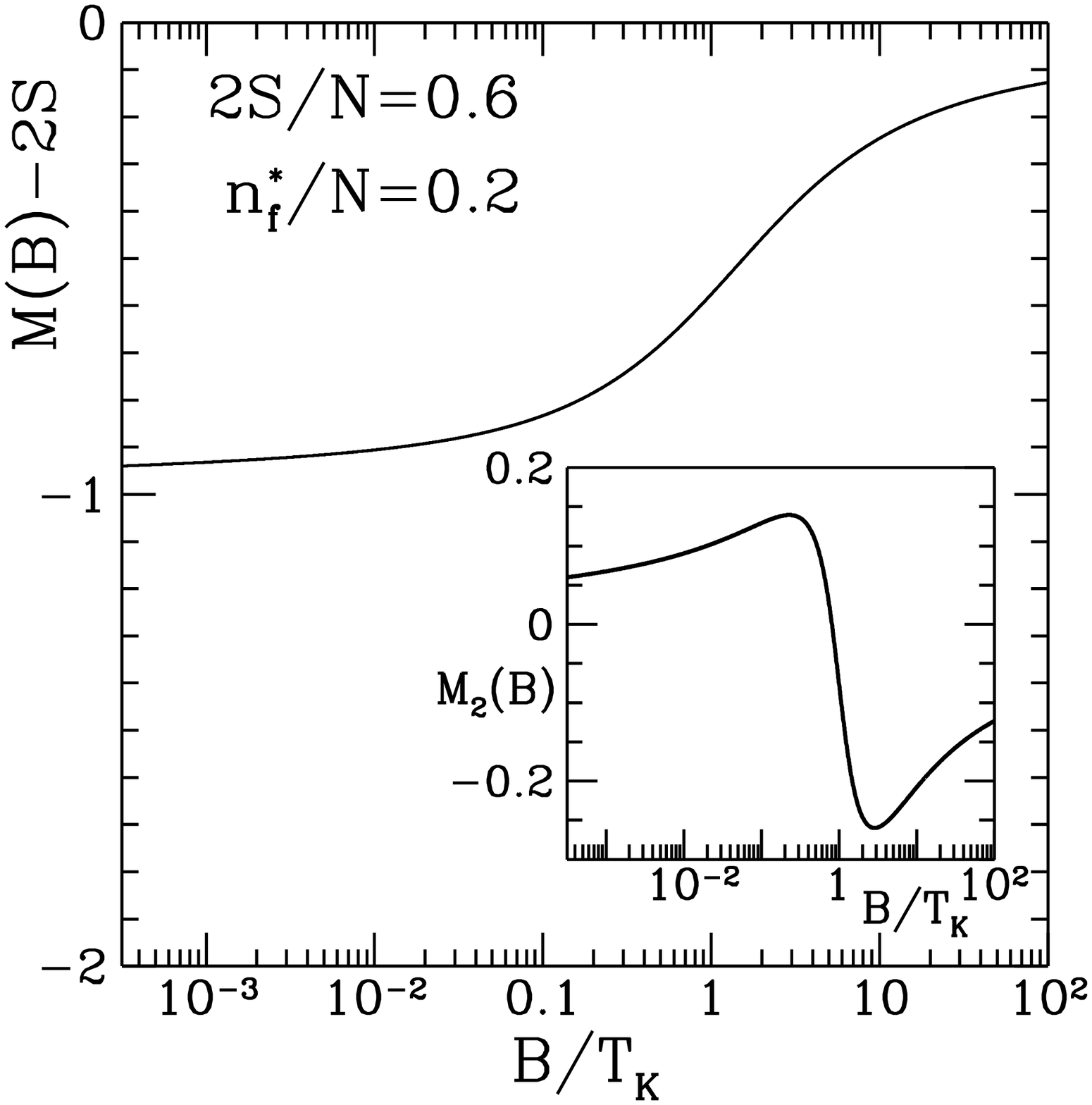}}\vskip 0.4truein
\caption{{Showing the field dependent magnetization
for a Kondo model with $\zeta >0$. In this case, the Kondo effect
is a single-stage process. Inset shows the corrections to the magnetization
derived from the residual interaction between the heavy fermi 
and spin fluid. These corrections are positive (ferromagnetic) at low
temperatures and remain small at al temperatures}}\label{fig2}\end{figure}}

We can check this result by completely removing
the fermionic contribution.  In the limit
$\Delta\rarrow 0$, $2 \zeta_o \rarrow T_K$, 
$M_1(B)$ and
$M_2(B)$ develop discontinuities which 
precisely cancel, to yield 
\bea
M(B)= 2S
-\frac{1}{\pi}\left\{
\arctan\left[\frac{ N \ln
\left[\frac{T_K}{B}\right]}{2S \pi}\right]+
\frac{\pi}{2}
\right\}
\eea
which is the residual magnetization of the
spin $S$ Kondo model. 

Let us now restore a finite $\Delta$. The first
term, $M_1(B)$ represents the screening of the local moment,
by the Kondo effect.
$M_1(B)$ has the limiting values
\bea
M_1(B)= \left\{
\begin{array}{lr}
2S-1&(B<< T_K)\cr
2S&(B>>T_K)
\end{array}
\right.
\eea
The second term $M_{2}(B)$ derives from the
screening effect produced by the residual interaction
between the Fermi liquid and the 
magnetic moment. 
For $\zeta>0$, this can
be re-written
\bea
M_2(B)= 
\frac{1}{\pi}\arctan\left[\frac{2S \pi}{
\frac{Q\Delta \pi}{2 \zeta_o -B} + N \ln
\left[\frac{T_K}{B}\right]}\right]
\eea
At low fields, this contribution is small
and positive, corresponding to an
irrelevant residual  ferromagnetic interaction.
At $2\zeta_o=B$, this contribution passes
continuously through zero, due to
a change of sign in the residual interaction. At
still higher fields, this correction remains
small, and asymptotes to zero.
\fg{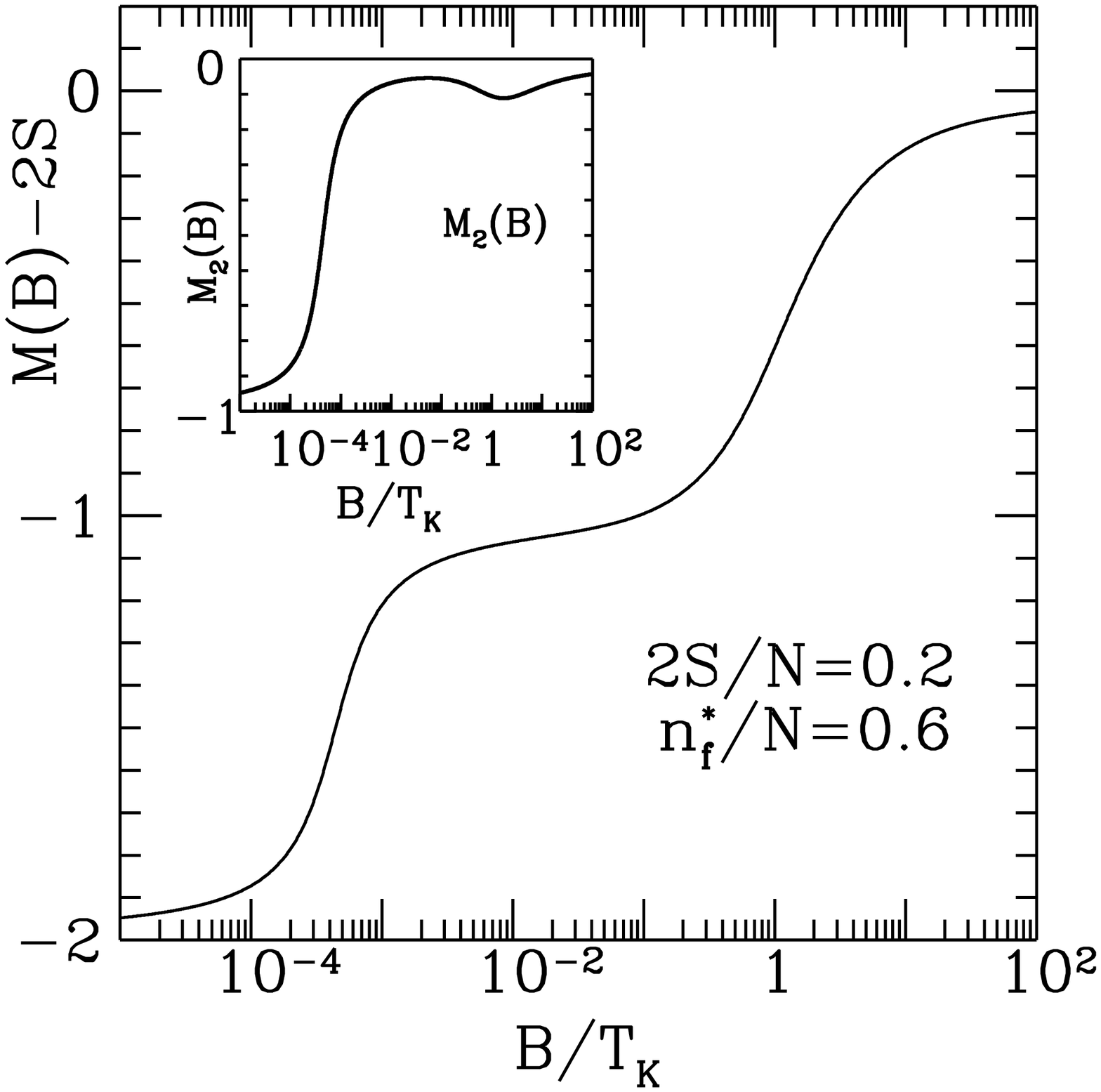}{
Field dependent magnetization
for a Kondo model with $\zeta <0$. In this case, the Kondo effect
is a two-stage process. The corrections to the magnetization
derived from the residual antiferromagnetic 
interaction between the heavy fermion 
and spin fluid.  These interactions grow to $-1$ at low temperatures, 
corresponding to a second screening process at the
renormalized Kondo temperature $T_K^*$.}{fig3}

By contrast, if $\zeta_o < B$, in this case the
residual interaction with the Fermi fluid is
{\sl antiferromagnetic}. Since $2\zeta_o-B$ is always
negative, we
can now write
\bea
M_2(B) = - \frac{1}{2} +
\frac{1}{\pi}\arctan
\left[
	\frac{
			\frac{
				Q\Delta \pi}{2 \zeta_o -B} 
			+ N \ln
			\left[\frac{B}{T_K}\right]}{2 \pi S}
\right]
\eea
For small fields, $M_2(B)\rarrow -1$, so that
this term constitutes an {\sl additional}
screening contribution to the magnetization.
For small negative zeta, we can approximate this
expression by
\bea
M_2(B) = - \frac{1}{2} -
\frac{1}{\pi}\arctan
\left[
	\frac{
			+ N \ln
			\left[\frac{T^*_K}{B}\right]}{2 \pi S}
\right]
\eea
where 
\bea
T_K^*= T_K e^{- \frac{\pi q \Delta }{|2 \zeta_o|}}= 
T_Ke^{- \pi q \cot [\pi(\tilde n_f^* - \frac{1}{2})]},
\qquad (n_f^*> \frac{1}{2})
\eea
which corresponds to a second screening process, governed by the
second-stage Kondo temperature $T_K^*$.

An alternative way to derive the same result is to consider how the
Schwinger boson field condenses in an applied field.
The constraint associated with the bosonic
part of the spin is written (\ref{newc})
\bea
2S &=& n_b +\frac{1}{Q}\theta\dg \theta
\eea
In zero field, the Fermi fluid is unpolarized, and
the magnetization is given by the condensed part
of the Schwinger Bose field. Suppose we apply a
small field that condenses the $b_1$ component,
so that
\bea
b_{\si} = \sqrt{M}\delta_{\si 1}+ \delta b_{\si}
\eea
then the constraint can be re-written as
\bea
2S = M + \sum_{\si}\la \delta b\dg _{\si}\delta b_{\si} \ra
+\frac{1}{Q}\la \theta\dg \theta\ra
\eea
or 
\bea
M = 2S -  \sum_{\si}\la \delta b\dg _{\si}\delta b_{\si} \ra
-\frac{1}{Q}\la \theta\dg \theta\ra
\eea
Diagrammatically, these two contributions to moment reduction are given by
\bea
\nonumber \\ 
\sum_{\si}\la \delta b\dg _{\si}\delta b_{\si} \ra
+\frac{1}{Q}\la \theta\dg \theta\ra=
\raisebox{-1cm}[0cm][0cm]{
\epsfig{file=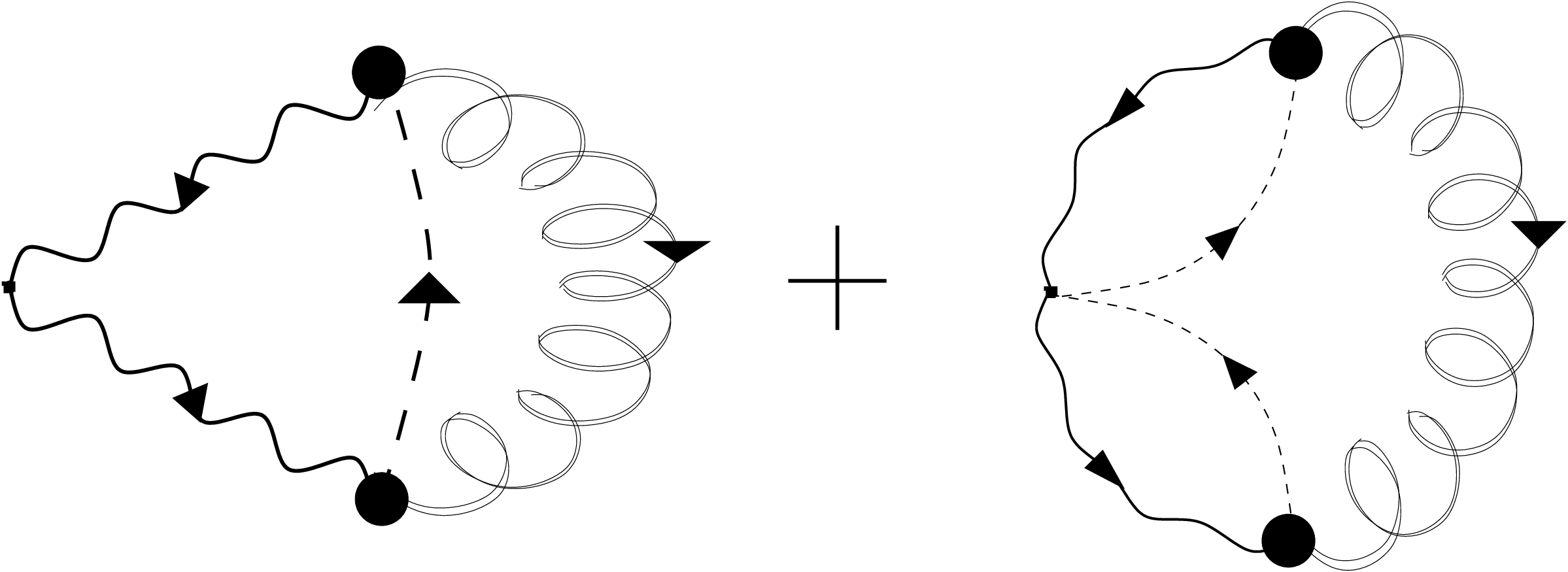,width=5cm}}
\\ \nonumber
\eea
Of course, only the combination of the two terms is  gauge
invariant, but by fixing the gauge, we can assign them each
physical meaning.
The first corresponds to fluctuations in
the direction of the local moment. The second represents the
reduction in the amplitude of the moment derived from the inter-conversion of spins into
heavy fermions. 
We can compute the sum of these diagrams by noting that they are generated
by differentiating the RPA diagrams
contributing to the fermionic zero-point energy with respect to the frequency:
\bea
\frac{\partial}{\partial i\omega_n}\left[\sum_{\rm loops}
\raisebox{-1.5cm}[0cm][1.5 cm]{
\epsfig{file=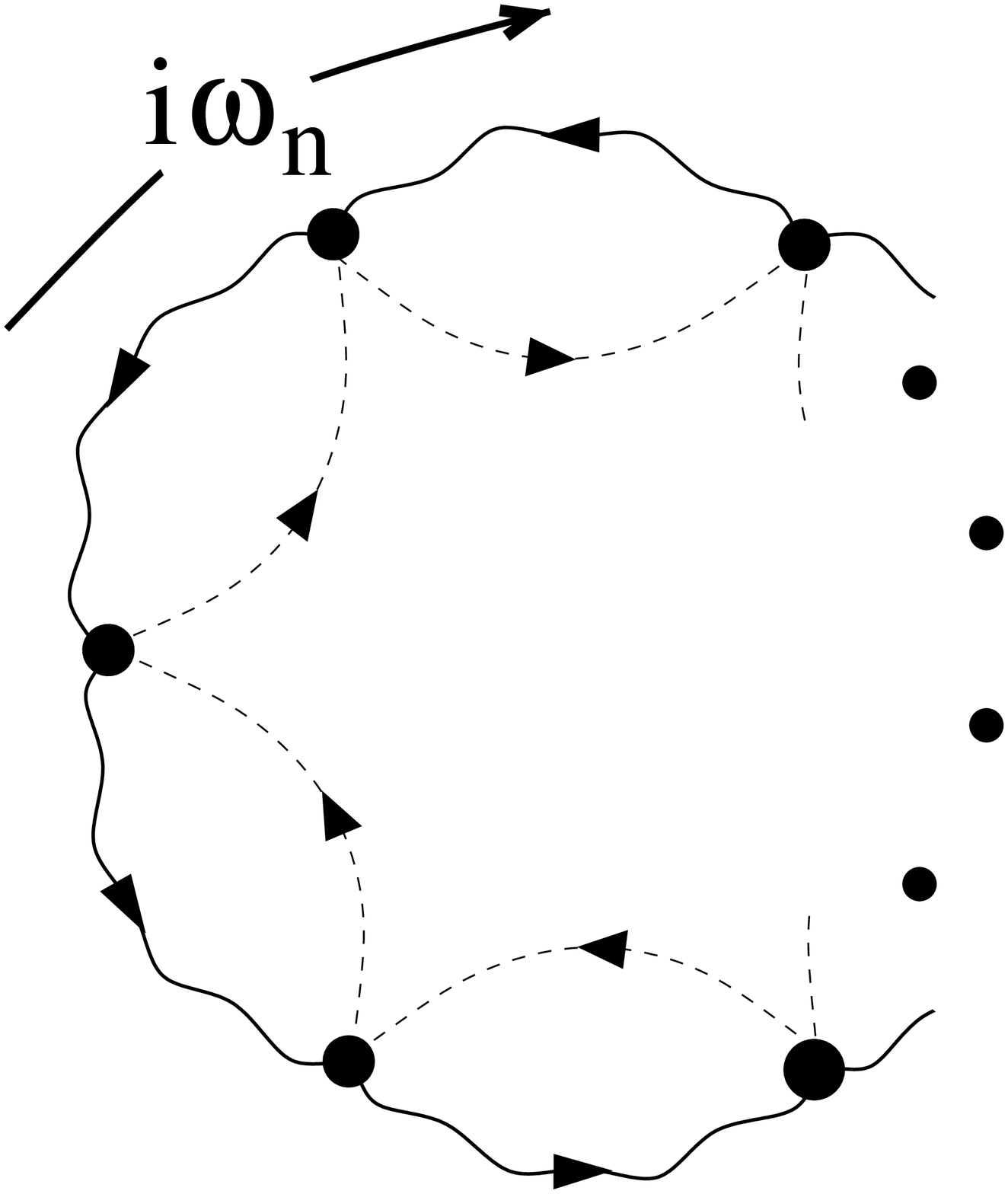,width=3cm}}
\right]= \sum_{\rm loops} \left[
\raisebox{-1.5cm}
[0cm][1.5 cm]{\epsfig{file=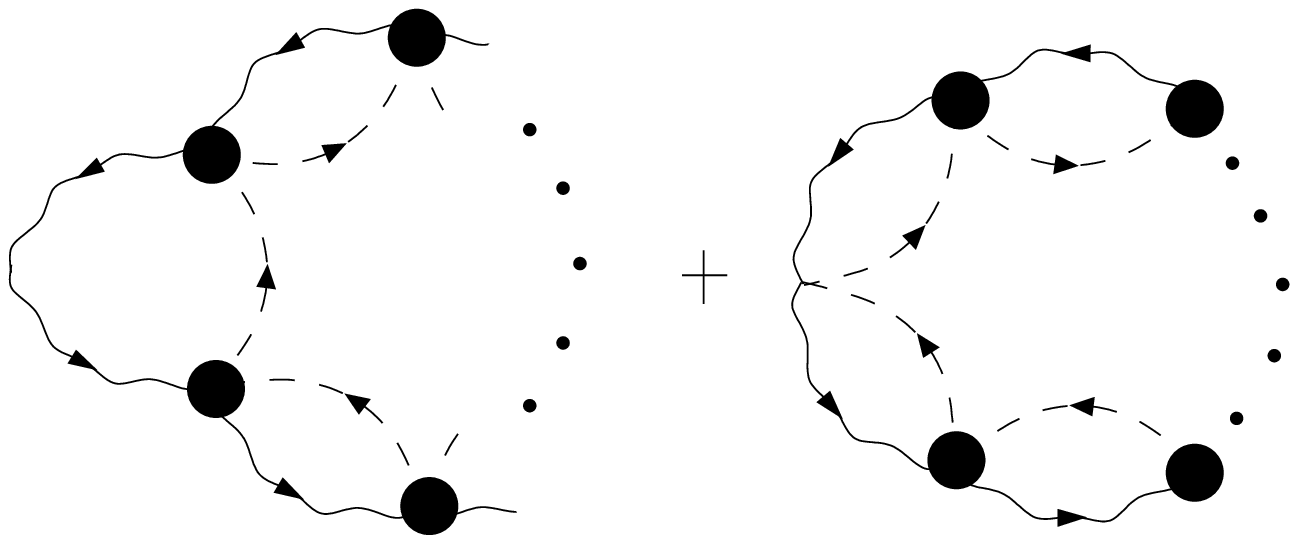,width=7cm}}\right]
\eea
so that 
\bea
\sum_{\si}\la \delta b\dg _{\si}\delta b_{\si} \ra
+\frac{1}{Q}\la \theta\dg \theta\ra=
- T \sum_{\omega= i \omega_n} \frac{\partial}{\partial \omega}
{\rm ln } [ P( \omega)]= - M_{\alpha}\eea
which enables us to identify the reduction in the magnetization with 
the fluctuations in direction and magnitude of the local moment. 
In this way, we see that the fluctuations which screen
the moment are given by
\bea
\sum_{\si}\la \delta b\dg _{\si}\delta b_{\si} \ra
+\frac{1}{Q}\la \theta\dg \theta\ra=
\left\{
\begin{array}{lr}
1, & (n_f^*<N/2)\cr
2, & (n_f^*>N/2)
\end{array}
\right.
\eea
depending on whether a one, or two-stage screening process takes place.
Although these results are only calculated to leading order in the
large-$N$ expansion, we expect the appearance of integer values for the screening is exact for a local moment. 

\section{Strong-coupling Picture of the Two-Stage Kondo effect}

To gain a complimentary
insight into the two-stage Kondo effect, it is useful to examine this
phenomenon 
in the strong-coupling limit.    Imagine a local moment, described by an L-shaped
representation of SU(N), denoted by  the Young Tableau
\setlength {\unitlength}{0.005\textwidth}
\bea
{\bf S}=\parbox{50\unitlength}{
\begin{picture}(50,40)(0,0)
\multiput(20,25)(5,0){5}{\framebox(5,5){}}
\put(28,37){\vector(-1,0){8}}
\put(37,37){\vector(1,0){8}}
\put(30,35){$2S$}
\multiput(20,20)(0,-5){4}{\framebox(5,5){}}
\put(13,22){\vector(0,1){8}}
\put(10,17){$n_f^*$}
\put(13,13){\vector(0,-1){8}}
\end{picture} }
\eea
In the ground-state of the strong-coupling Hamiltonian
\bea
H_K= \frac{J}{N}c\dg_{0\alpha}\pmb{$\Gamma$}_{\alpha \beta}c_{0\beta} \cdot
{\bf S}
\eea
electrons form a singlet with the fermionic part of the spin creating a partially screened
moment,  denoted by a Young-tableau with a completely filled first row.
\bea
{\bf S}^*=(\pmb{$\Gamma$}_{e_0}+{\bf S})=\parbox{50\unitlength}{
\begin{picture}(50,60)(0,0)
\multiput(20,45)(5,0){5}{\framebox(5,5){}}
\put(28,57){\vector(-1,0){8}}
\put(37,57){\vector(1,0){8}}
\put(30,55){$2S$}
\multiput(20,40)(0,-5){7}{\framebox(5,5){}}
\put(13,35){\vector(0,1){15}}
\put(10,30){$N$}
\put(13,26){\vector(0,-1){15}}
\put(20,12){$c_0$}
\put(20,17){$c_0$}
\put(20,22){$c_0$}
\end{picture} }
\equiv\parbox{50\unitlength}{
\begin{picture}(50,15)(0,0)
\multiput(20,5)(5,0){4}{\framebox(5,5){}}
\put(27,17){\vector(-1,0){7}}
\put(26,17){\vector(1,0){14}}
\put(22,23){$2S-1$}
\end{picture}
}
\eea
where in this example we have taken $N=8$.
Since the first column of the tableau is a singlet (with $N$ boxes), it 
can be removed from the tableau,leaving behind a partially screened
spin $S-1/2$, described by a row Young-tableau with $2S-1$ boxes.
If we now couple the electron at the origin with electrons at site `1'
via a small hopping matrix element $t<<J$, then the virtual charge fluctuations
of electrons in and out of the singlet at the origin will lead to a residual
coupling
between the partially screened moment and the electrons
at the neighboring site '1'
\bea
H_{(1)}= \frac{J^* }{N}{\bf S}^*\cdot c\dg_{1\alpha} \pmb{$\Gamma$}_{\alpha \beta}
c_{1\beta}
\eea
where $J^*\sim t^2/J$. 
In the  SU(2) Kondo model, only electrons  parallel
to the residual moment ${\bf S}^*$ can hop onto the origin, which gives
rise to a ferromagnetic coupling $J^*<0$.  In the 
SU(N) case, electrons can hop  provided they are not in the same
spin state as electrons  at the origin.
The sign of the coupling $J^*$  depends on the number of conduction
electrons $n_c= n-n_f^*$, bound at the origin. If the $n_c=N-1$, 
electrons hopping onto the origin will have to be parallel to the residual
spin, so in this case the coupling is ferromagnetic, $J^*<0$. By contrast,
if $n_c<<N$, there are many ways
for the electron to hop onto the origin with a spin component that is
different to the residual moment, so the residual
interaction will be antiferromagnetic, $J^*>0$.  By carrying out a large
$N$ calculation in the strong coupling limit or by making a 
detailed strong coupling calculation for SU(N), we are able to confirm that
for $N>2$, 
$J^*$ changes sign when the number of bound-conduction electrons is less than
$N/2$, and in the large $N$ limit, it is given by\cite{pepin2000}
\bea
J^*= - \frac{t^2}{J (1 -\tilde n_f^*)\tilde n_f^*}
\left[\frac{\frac{1}{2} - \tilde n_f^*}{(1 -
\tilde n_f^* + \tilde n_b)}\right]
\eea
where $\tilde n_f^*= n_f^*/N$ and $\tilde n_b = 2S/ N$. 

When  $n_f^*>N/2$, $4J^*>0$, 
the strong-coupling fixed point becomes unstable, and a second-stage Kondo effect
occurs, binding a further 
$N-1$ electrons at site "1" to form a state
denoted by the tableau
\begin{equation}
{\bf S}^{**}=(\pmb{$\Gamma$}_{e_0}+\pmb{$\Gamma$}_{e_1}+
{\bf S}=\parbox{50\unitlength}{
\begin{picture}(50,60)(0,0)
\multiput(20,45)(5,0){5}{\framebox(5,5){}}
\put(28,57){\vector(-1,0){8}}
\put(37,57){\vector(1,0){8}}
\put(30,55){$2S$}
\multiput(20,40)(0,-5){7}{\framebox(5,5){}}
\put(13,35){\vector(0,1){15}}
\put(10,30){$N$}
\put(13,26){\vector(0,-1){15}}
\put(21,12){$c_0$}
\put(21,17){$c_0$}
\put(21,22){$c_0$}
\multiput(25,40)(0,-5){7}{\framebox(5,5){}}
\put(26,12){$c_1$}
\put(26,17){$c_1$}
\put(26,22){$c_1$}
\put(26,27){$c_1$}
\put(26,32){$c_1$}
\put(26,37){$c_1$}
\put(26,42){$c_1$}
\end{picture}}
\equiv
\parbox{50\unitlength}{
\begin{picture}(50,15)(0,0)
\multiput(20,5)(5,0){3}{\framebox(5,5){}}
\put(28,15){\vector(-1,0){8}}
\put(25,15){\vector(1,0){10}}
\put(20,20){$2S-2$}
\end{picture}
}
\end{equation}
corresponding to a residual spin $S^{**}=S-1$, $M=2S-2$.
This final configuration is stable, because  an electron at site
"2" can only hop onto site "1"  if it is parallel to the
unquenched moment. 

We see that our supersymmetric approach permits us to examine the consequences of this
two-stage Kondo effect, starting from weak coupling. Translated into the
weak coupling language, 
the two vertical columns of the Young tableau will correspond to
two separate screening clouds, of very different radii
\bea
l= \frac{v_F}{T_K}, \qquad \hbox{and}\qquad l^*= \frac{v_F}{T_K^*},
\eea
respectively.   It is remarkable that  a
point-like complex impurity can give rise to two separate length-scales in this way.

\section{Discussion}

In this paper we have developed  a  spin representation that
interpolates between the Schwinger boson and Abrikosov pseudo-fermion representations, which exhibits the property of supersymmetry. 
As an exploratory exercise, we have applied the method to 
to a class of single impurity Kondo models, where 
we have been able to
examine how local moment behavior emerges as the strength of the Hund's interactions between
the spins is 
systematically increased.  
Suppose we consider a spin representation with $Q$ boxes,
and examine how the ground-state evolves as we progressively
increase the moment from $S=1/2$ to $S=Q/2$. 
One of the surprising discoveries, is that there
are in fact, two routes by which the magnetic moment emerges from the
Fermi liquid (Fig. 5): 

\begin{enumerate}

\item{One stage Kondo effect, where $Q< N/2$. 
Once the spin $S$ exceeds one half, a partially
screened moment is generated.  The 
low temperature fixed point is described by a co-existence of a Fermi liquid and
a  moment of spin $S^*=S-1/2$, with a slowly vanishing ferromagnetic coupling
between the two degrees of freedom.  }

\item{ Two stage Kondo effect, where 
$Q>N/2$. 
At intermediate scales, the moment quenches to a spin $S^*=S-1/2$, but the
residual coupling to the conduction sea is now antiferromagnetic, and
a second-state quenching occurs to a spin $S^{**}=S-1$.   
When 
the starting spin is $S=1$, 
a new singlet phase is formed, with 
one additional fermionic bound-state. ( We label this phase $FL^*$ in
Fig. 5, but do not at this stage know if this state is a Fermi liquid.)
 }

\end{enumerate}
\begin{figure}[tb]
\epsfxsize= 0.6 \textwidth
% ***********For one column  ********************
%\epsfxsize=7.0in 
% ***********************************8
\centerline{\epsfbox{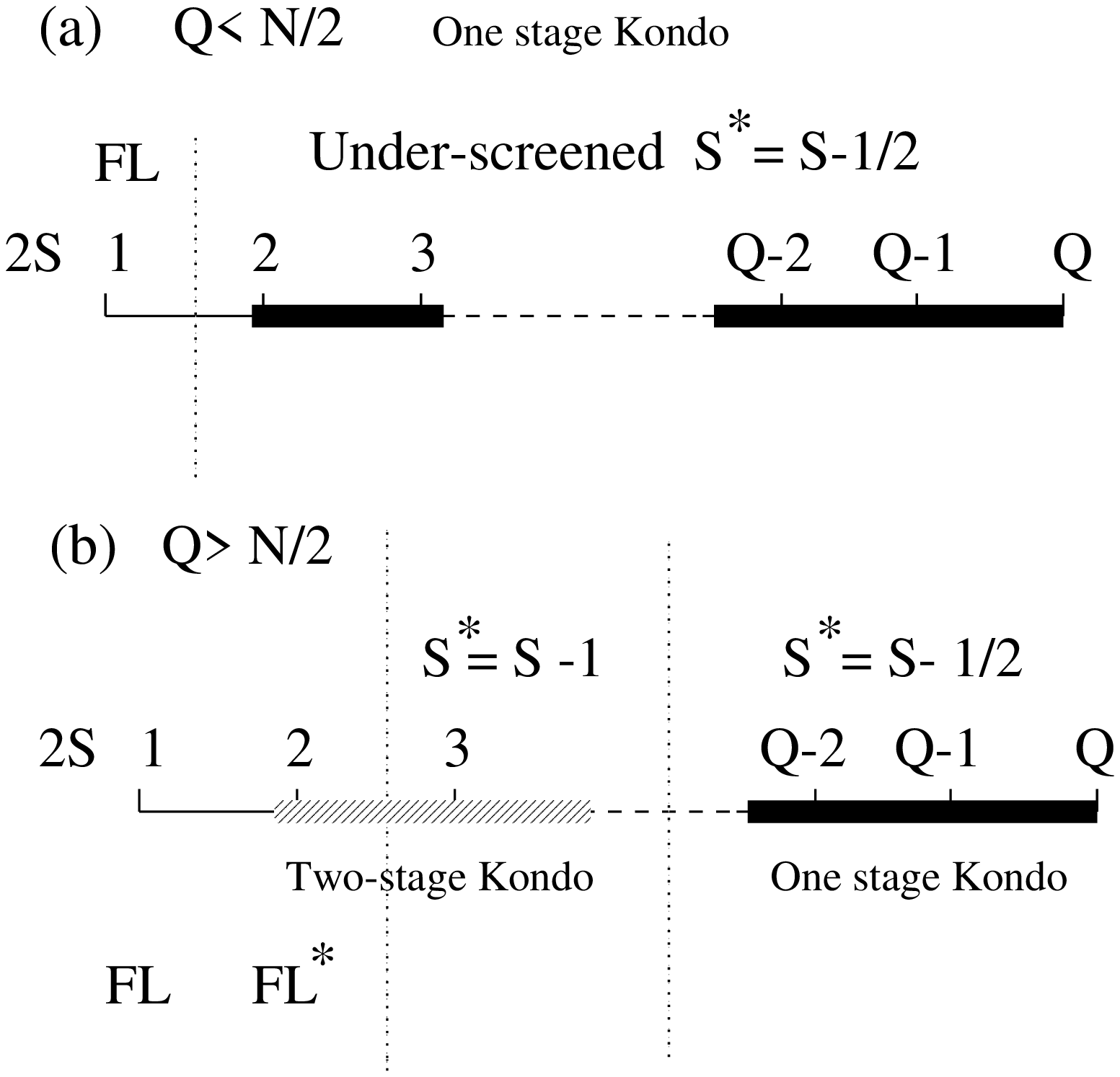}}
\vskip 0.1truein
\protect\caption{
Two scenarios for the emergence of magnetism as the size of $S$ is
progressively increased. (a) Single Stage Kondo effect, where  for
$S> \frac{1}{2}$,  a partially quenched moment with spin $S^*=S=\frac{1}{2}$.
(b)For $Q>N/2$, the initial emergence of a local moment
is accompanied by a two-stage Kondo effect, where, provided $n_f^*>N/2$, 
the spin is screened from $S$ to $S^**= S-1$. 
}
\label{fig5}
\end{figure}

We should now like to discuss the future
extension of this approach to a Kondo lattice. 
Two decades ago, Doniach \cite{doniach77}
argued that 
the properties of a Kondo lattice should depend critically on the ratio  of
the Kondo temperature to the RKKY interaction $\kappa= T_K/J_{RKKY}$.
Heuristically,  the Kondo and RKKY scales are related to the Kondo coupling
constant according to 
\bea
T_K\sim D e^{ - 1/ J \rho} , \qquad J_{RKKY} \sim J^2 / D
\eea
where $D$ is the conduction band-width. Doniach pointed out that
$\kappa$ grows with the size of $J$, arguing that 
for small $J$, 
the system is expected to be antiferromagnetically ordered, but for large
$J$, the magnetism melts to form a ``heavy''
Fermi liquid. Unfortunately, our theoretical
understanding is at present, only limited to a
discussion of energy scales, and little is known
about the nature of the transition between these
two limiting cases. 

Can we do to shed light on these issues by extending the supersymmetric
approach to the lattice?  
The mean-field solution will in general give rise to 
a heavy Fermi liquid which co-exists with a
lattice of under-screened moments. The 
Fermi surface volume $V_{FS}$ 
will depend on the size $M$ of the under-screened moments, 
\bea
\frac{V_{FS}}{(2 \pi)^3} = (n_c + n_f^*)/N= (n_c+Q-M)/N
\eea
At the 
mean-field level, these moments can point in any direction, but  once we
include the effects of the Gaussian fluctuations of the $\alpha$ fermions,
two effects will take place:
\begin{itemize}

\item The local moments will be partially screened by the Kondo
effect with the heavy electron fluid.

\item The fluctuation free energy will be sensitive to the direction
in which the spins condense.

\end{itemize}
Typically, we expect  the fluctuation free energy will be lowest in an
antiferromagnetic spin configuration, where, for instance
\bea
\la b_{\si}({\bf x})\ra = \sqrt{2 M} \bigl[ \cos^2 ( {\bf Q}\cdot {\bf x}/2)
\delta_{\si 1} + \sin^2 ( {\bf Q}\cdot {\bf x}/2)
\delta_{\si 2}\bigr]
\eea
This dependence on the relative orientation of the
moments defines  a renormalized ``RKKY''
interaction
$J_{RKKY}^*$. By tuning $J$ we will be able to examine how the RKKY
interaction is renormalized by the Kondo effect and examine how the
staggered magnetization depends on the screening process
\bea
M&=& 2S - m(J)\cr
m(J) &=&  \sum_{\si}\la \delta b\dg _{\si}\delta b_{\si} \ra
+\frac{1}{Q}\la \theta\dg \theta\ra
\eea
where $m(J)$ is a continuous function of $J$. The critical
value where $M$ vanishes (at small $S$)  defines the point where the
magnetism is  eradicated by the Kondo effect. In this way, we hope
to be able to carry a model calculation of the phase diagram
first envisaged by Doniach. 

An open question, is whether this new approach can shed light on the
nature of the quantum critical point separating the magnetic and the paramagnetic
phases of the Kondo lattice.   Ultimately we are interested in the
properties of a Kondo lattice of elementary spins corresponding
to one box in Young Tableau. 
This large $N$ approach can provide information about
the properties of a class of $L$-shaped representations. If there
is any universality associated with the emergence of magnetism
at absolute zero, then perhaps a large
$N$ approach will enable us to triangulate on the properties of a real
Kondo lattice. (Fig. 6)
\begin{figure}[tb]
\epsfxsize= 0.6 \textwidth
% ***********For one column  ********************
%\epsfxsize=7.0in 
% ***********************************8
\centerline{\epsfbox{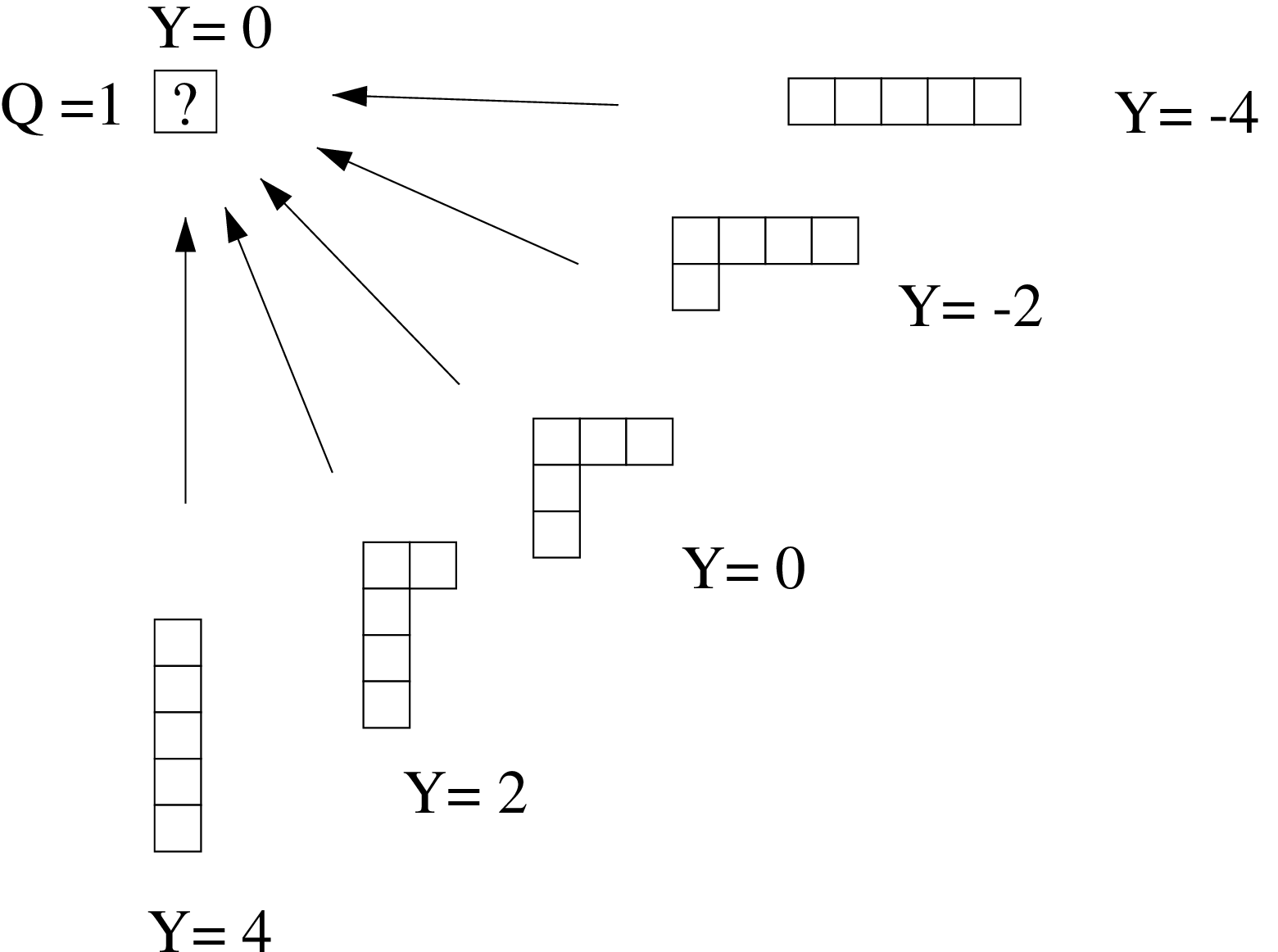}}
\vskip 0.1truein
\protect\caption{
Illustrating the idea that the properties of a class of $L$ shaped
representations will enable us to triangulate on the properties
of a small $N$ Kondo lattice. 
}
\label{fig6}
\end{figure}

One of the interesting aspects of the supersymmetric approach, is the appearance
of  fermionic ``phase fluctuations'' between the
spin and the heavy electron fluid: it is these fluctuations, described
by the gauge fermion $\alpha$, 
which  mediate the interaction between the magnetic condensate and the heavy
electron fluid.
 In the case where the ``mass'' of this excitation,
$\zeta$ is positive, the gauge fermion $\alpha$ can be integrated out of
the problem, and the interaction between the magnetic and electron fluid
could be treated as a point interaction. However, when $\zeta$ is negative,
the
$\alpha$ fermion gives rise to a new bound-state.  Will the same phenomenon
occur in the lattice? This  could lead to the possibility of two different
kinds of fixed point:
\begin{itemize}

\item A Millis Hertz fixed point\cite{hertz,millisqcp}, where the weak
ferromagnetic interaction between the magnetic and electron fluid can be
treated as a point vertex. In this situation, the transition will be
described by the interaction between a Gaussian magnetic fluid and a
well-defined Fermi surface.

\item A non-Fermi liquid fixed point, where the dynamical fermion mediating
the magnetic interaction becomes an active participant in the physics.  
If the gauge fermion
developed gapless excitations, then the decay of heavy
fermions into unquenched spins, described by the process
\bea
f_{\si}^- \rightleftharpoons b_{\si} + \alpha^-,
\eea
would lead to a phase with a  novel kind of spin-charge separation. 

\end{itemize}
These points will be examined in greater detail in a future publication.

We should like to thank Natan Andrei, Antoine Georges, Mireille Lavagna,
Olivier Parcollet and Revaz  Ramazashvili for critical discussions related
to this work.  This research was supported in part by NSF grant DMR 961999
and research funds from the EPSRC, UK.

\vskip 0.2 truein
\noindent{\bf Appendix A: Operator Expression for the Cazimir }

In this section, we prove that the Cazimir 
\bea
{\bf S }^2 = ( b\dg\Gamma b + 
f\dg\Gamma f)^2
\eea
can be written 
\bea
{\bf S}^2 =  Q ( N - \hat{\cal Y}- \frac{Q}{N})\label{cazimir2}.
\eea
To show this relationship, we use the completeness result. Using the
normalization
\bea
{\rm Tr} [ \Gamma^{a} \Gamma^b]= \delta^{ab}
\eea
this is
\bea
\Gamma^{\lambda}_{ab}\Gamma^{\lambda}_{cd} + \frac{1}{N}
\delta_{ab} \delta _{cd} = \delta_{ad} \delta_{bc}
\eea
By using this to expand
\bea
 {\bf S}^2
= ( b\dg\Gamma b + 
f\dg\Gamma f)^2
\eea
we obtain
\bea
 {\bf S}^2
 = - \frac{1}{N} Q^2 +
( \bs{a}\dg\bs{b} + \fs{a}\dg\fs{b}) 
( \bs{b}\dg\bs{a} + \fs{b}\dg\fs{a}) 
\eea
We can expand each term in this expansion as follows:
\bea
 \bs{a}\dg\bs{b}\bs{b}\dg\bs{a} 
+ \fs{a}\dg\fs{b}\fs{b}\dg\fs{a} &=&  n_b( n_b + N-1)-n_f( n_f - (N+1))\cr
&=& n_b^2 - n_f^2 + N Q + (n_f - n_b)
\eea
Also
\bea
\fs{a}\dg\fs{b} \bs{b}\dg\bs{a}= - \bs{b}\dg\fs{b} \fs{a}\dg\bs{a}+ n_b
\eea
Combining these results we obtain
\bea
{\bf S}^2&=& -\frac{1}{N} Q^2 + ( n_b^2 - n_f^2)  +  ({N+1})Q
- 2
\bs{b}\dg\fs{b} \fs{a}\dg\bs{a}\cr
&=& 
-\frac{1}{N} Q^2  +  ({N+1})Q 
+  Q [ n_b - n_f - \frac{2}{Q} \theta \theta\dg]\cr
&=&-\frac{1}{N} Q^2  + {N}Q 
-  {Q} [ n_f - n_b+ \frac{1}{Q}[ \theta ,\theta\dg] ]\cr
&=&  Q ( N - \hat{\cal Y}- \frac{Q}{N})
\eea
where we have used the result $2\theta \theta\dg =  [ \theta ,\theta\dg ]+ Q$
to carry out the last step but one. 

\noindent{\bf Appendix B: Supersymmetry of Lagrangian}
\vskip 0.2 truein

In this section, we examine the transformation of the Lagrangian 
\bea
\Ls^*  &=&  \sum_{\si}\Psi\dg_{\si}\biggl(
\partial_{\tau} +\lambda+\mat{\zeta &
\slp \alpha\cr 
\bar \slp\alpha \dg &
-\zeta}
 \biggr)
\Psi_{\si} +
Q_o\alpha \dg\slp \alpha
\eea
under the transformation $
\Ps\rarrow h \Ps$ , where  
\bea
h
= e^{i \theta_Q}
\mat{\sqrt{1 - \eta \bar \eta} & \eta\cr -\bar \eta & \sqrt{1- \bar \eta
\eta}}\mat{e^{i\theta_{\zeta}}& 0 \cr
0 & e^{-i\theta_{\zeta}}},\eea
is a general member of the group $SU(1|1)$. Under this transformation, 
\bea
\Ls^*  \rarrow \Ls^* + \sum_{\si}\Psi\dg_{\si}\biggl(
(h\dg\partial_{\tau}h) +h\dg[\mat{\zeta &
\slp \alpha\cr 
\bar \slp\alpha \dg &
-\zeta},h]
 \biggr)
\Psi_{\si}
\eea
When we expand the correction, we obtain
\bea
\sum_{\si}\Ps\dg(h\dg \D h) \Ps = \sum_{\si}\Ps \dg
\biggl[
i \dot \theta_Q + i \dot \theta_{\zeta}\tau_3 + \bar \eta \D \eta 
+
\mat{ & e^{-2 i \theta_{\zeta}}\D \eta\cr - e^{2 i
\theta_{\zeta}}\D \bar \eta & }
\biggr]\Ps,\label{1st}
\eea
\bea
\sum_{\si}\Ps\dg  h\dg[\mat{\zeta &
\cr 
 &
-\zeta},h]\Ps=\sum_{\si}\Ps\dg \mat{
&  2 \zeta \eta e^{-2 i \theta_{\zeta}}\cr
2 \zeta \bar\eta  e^{2 i
\theta_{\zeta}}&
}\Ps +  Q ( 2 \zeta \bar \eta \eta) \label{2nd}
\eea
and
\bea
\sum_{\si}\Ps\dg  h\dg[\mat{ &
\slp \alpha\cr 
\bar \slp\alpha \dg &
},h]\Ps=   \sum_{\si}\Ps\dg \mat{ \bar \eta \slp \alpha +
(\bar \slp  \alpha \dg) \eta & (e^{-2 i
\theta_{\zeta}}-1)\slp \alpha\cr  (e^{2 i \theta_{\zeta}}-1)\bar
\slp\alpha \dg & \bar \eta \slp \alpha +
(\bar \slp  \alpha \dg) \eta}\Ps\label{3rd}
\eea
Combining (\ref{1st}) , (\ref{2nd}) and (\ref{3rd})we obtain
\bea
\Ls^* \rarrow  \sum_{\si}\Psi\dg_{\si}\biggl(
\partial_{\tau} +\lambda'+\mat{\zeta' &
\slp' \alpha'\cr 
\bar \slp'\alpha ^{'\dagger} &
-\zeta'}
 \biggr)
\Psi_{\si} +
Q_o\alpha ^{'\dagger}\slp '\alpha '
\eea
where $\lambda' = \lambda + i \dot \theta_Q$, $\zeta'=\zeta + i \dot
\theta_{\zeta}$ , $\alpha' = e^{-i \theta_{\zeta}}(\alpha +
\eta)$,$\alpha^{'\dagger} = e^{i \theta_{\zeta}}(\alpha\dg +
\bar \eta)$. The primes on the gauged derivatives denote
$\slp ' = (\partial + 2 \zeta')$ and  $\bar \slp ' = (-\partial +
2
\zeta')$. The Lagrangian
is thus gauge invariant under the transformation
\bea
\Ps&\rarrow &h \Ps
, {\cal V} \rarrow h {\cal V},\cr
\lambda &\rarrow &\lambda - i \dot \theta_Q, \ 
\zeta \rarrow \zeta - i \dot \theta_{\zeta},\cr
\alpha &\rarrow &e^{2 i \theta_{\zeta}}\alpha -
\eta, \ 
\alpha\dg \rarrow e^{-2 i
\theta_{\zeta}}\alpha\dg - \bar \eta.  
\label{fullg2}
\eea
\vfill\eject
\noindent{\bf Appendix C: Evaluation of fermionic Mean-Field Free
energy }
\vskip 0.2 truein

We wish to calculate
\bea
F_{f} = - NT \sum_n \ln [ \lambda_f +i \Delta_n -i \omega_n]
e^{i \omega_n 0^+}
\eea
where $\Delta_n=- \Delta {\rm sign }\omega_n$.
We shall regulate this sum by calculating 
\bea
F_{f} = - NT \sum_n\biggl( \ln [ \lambda_f +i \Delta_n-i \omega_n] - [ \lambda_f \rarrow \lambda _f + D]\biggr)e^{i \omega_n 0^+}
\eea
which we re-write as
\bea
F_{f} &=& - 2NT {\rm Re}\sum_{n\ge 0}\bigl(\ln[ \xi + i \omega_n]- \ln[ \xi + D +i \omega_n]\bigr)e^{i n 0^+} 
\cr
&=& -2 NT {\rm Re}\sum_{n\ge 0}\biggl(\ln\biggl[ \frac{\xi}{2 \pi i T}+ \frac{1}{2} +n\biggl]-
\ln\biggl[ \frac{\xi+D}{2 \pi i T}+ \frac{1}{2} +n\biggl]
\biggr) e^{i n 0^+} 
\eea
where $\xi = \lambda_f+ i \Delta$.  Next,
using the result
\bea
\sum_{n\ge 0}\bigl(\ln[ b+n]- \ln[a+n]\bigr)e^{i n 0^+}=
\ln\left(\frac{\Gamma[a]}{\Gamma[b]}\right) + \frac{i \pi}{2}(b-a)
\eea
this becomes
\bea
\frac{F_{f}}{N}= - 2 T {\rm Re} \ln\left(\frac{\Gamma[\frac{\xi+D}{2 \pi i T}+\frac{1}{2}]}{\Gamma[\frac{\xi}{2 \pi i T}+\frac{1}{2}]}\right) - \frac{D}{2}
\eea

\vskip 0.2truein
\noindent{\bf Appendix D: Calculation of $P(
\omega)$} 
\vskip 0.2truein

We begin by writing
\bea
P(\omega) = 1 + \left(
\frac{\omega-2 \zeta}{q}\right)\Phi(\omega)
\eea where
\bea
\Phi(\omega) = \frac{T}{N}
 \sum_{ \si \nu}
G_{f \si}( \omega+ \nu)G_{b\si} ( \nu)
\eea
To calculate
$\Phi(\omega)$, we replace the discrete Matsubara
sum by  a Contour integral, to obtain
\bea
\Phi(\omega) = \frac{1}{N}\sum_{\si} \int \frac{ dz}{2 \pi i} n(z)
G_{b \si}( z)G_{f\si }( z + \omega)
\eea
where the integral runs counter-clockwise  around the poles in the 
Green's functions. 
Using the spectral decomposition, 
\bea
G_{f\si}(z) = \int \frac{ d \omega}{\pi} A_{f \si } ( \epsilon)
\frac{1}{z- \epsilon}
\eea
this becomes
\bea
\Phi(\omega) = -\frac{1}{N}\sum_{\si}
 \int \frac{ d \epsilon}{\pi}
[ n_{b \si}+ f( \epsilon) ] 
\frac{1}{\omega - \epsilon+ \lambda_{b\si}} A_{f \si } ( \epsilon)
\eea
where $n_{b \si} = n( \lambda_{b \si})$ is the Bose occupancy.
Now 
\bea
\sum_{\si} \int \frac{ d \epsilon}{\pi}
\frac{n_{b \si}}{\omega - \epsilon+ \lambda_{b\si}} A_{f \si } ( \epsilon)
&=
& \sum_{\si} \frac{n_{b \si}}{\omega - ( \lambda_{f \si}- \lambda_{b \si}) + i \Delta_n} \cr
&=
& \frac{2S}{\omega - 2 \zeta + i
\Delta_n}= 2S G_f(\omega+ B)
\eea
where we have replaced $2 \zeta
\rarrow 2 \zeta_o-B$, and $G_f(i\om_n)= (i\om_n -
2\zeta_o+ i
\Delta_n)^{-1}$ is the f-propagator in the absence
of a field, 
so that
\bea
P ( \omega) = 1 - \frac{ \omega + B - 2
\zeta_o}{q}\left[ G_f( \omega+B)\tilde n_b + I
\right]
\eea
where $\tilde n_b=  2S/N$ and 
\bea
I&=& -\frac{1}{N}\sum_{\si}
 \int \frac{ d \epsilon}{\pi}
f( \epsilon)  
\frac{1}{\omega - \epsilon+ \lambda_{b\si}} A_{f \si } (
\epsilon)
\cr&=& \frac{1}{N}\sum_{\si} \int
 \frac{ d \epsilon}{\pi}
f( \epsilon+m_{\si}B) 
\frac{1}{\omega - \epsilon+ \lambda_b} A_f (
\epsilon)
\eea
where $A_f(\omega)= Im G_f( \omega- i \delta)$.
Now since $N-2$ of the levels are unshifted,
to leading  order in the large $N$ expansion, 
we can set $m_{\si}=0$ in this expression. 
Also, since $\lambda_b=B$ in a magnetic field,
we can write $I= I(\omega+B)$ where, at $T=0$, 
\bea
I(\omega) &=& \int_{-D}^0 d \epsilon \frac{ A_f(
\epsilon)}{\pi} \frac{1}{\om - \eps}
\eea
Combining these results together, we can write
\bea
P ( \omega,B) =P_o(\omega+B)
\eea
where
\bea
P_o(\omega) =  1 - \frac{ \omega  -
2
\zeta_o}{q}\left[ G_f( \omega)\tilde n_b +
I(\omega)
\right]
\eea
is the zero field form of $P(\omega)$. 
Going on to 
evaluate  $I(\omega)$, we obtain
\bea
I(\omega) &=& \int_{-D}^0 d \epsilon \frac{ A_f(
\epsilon)}{\pi} \frac{1}{\om - \eps}
\cr
 &=&\int_{-D}^0 \frac{d \epsilon}{2 \pi i} 
\left(\frac{1}{\epsilon - \xi} 
-
\frac{1}{\epsilon - \xi^*} 
\right)\frac{1}{\om - \eps}\cr
&=& - \int_{-D}^0 \frac{d \epsilon}{2 \pi i} 
\left\{\left(\frac{1}{\epsilon - \xi} 
-
\frac{1}{\epsilon - \omega} 
\right)\frac{1}{\xi - \om} - [ \xi \rarrow \xi^*]
\right\}\cr
&=&  \frac{1}{2 \pi i} 
\left\{
\frac{1}{\om -\xi }\ln \left( {\xi \over \om}\right) 
 - 
\frac{1}{\om -\xi^* }\ln \left( {\xi^* \over \om}\right) 
\right\},
\eea
so that 
\bea
P_o ( \omega) = 
1 - \frac{( \om - 2 \zeta_o)}{q}\left[
\frac{ n_b } { \om - \xi} +  \frac{1}{2 \pi i} 
\left\{
\frac{1}{\om -\xi }\ln \left( {\xi \over \om}\right) 
 - 
\frac{1}{\om -\xi^* }\ln \left( {\xi^* \over \om}\right) 
\right\}
\right]
\eea
Now by writing $\xi = T_K e^{i \pi \tilde n_f}=
 \xi^* e^{i 2\pi \tilde n_f}$, where $\tilde
n_f= n_f^* /N$,  we can put this in the form
\bea
P_o(\omega) = 
1 - \frac{( \om - 2 \zeta)}{q}\left[
\frac{ \tilde n_b+ \tilde n_f } { \om - \xi} +  \frac{1}{2 \pi i} 
\ln \left( {\xi^* \over \om}\right) 
\left\{
\frac{1}{\om -\xi }
 - 
\frac{1}{\om -\xi^* }
\right\}
\right]
\eea
Since $\tilde n_b + \tilde n_f= q$, there are cancellations between
the first two terms which give
\bea
P_o( \omega) = \left[
\left(\frac{- i \Delta}{\omega- \xi}
\right)
-  \frac{( \om - 2 \zeta)}{\pi q}A_f( \omega) \ln \left( {\xi^* \over \om}\right)
\right]
\eea
Another useful way to rewrite this expression, 
is
\bea
P_o( \omega)= \left[
\left(\frac{- i \Delta}{\omega- \xi}
\right)
-  \frac{\Delta}{q \pi} { \rm Re}\left( \frac{1}{\om - \xi}\right) \ln \left( {\xi^* \over \om}\right)
\right]
\eea
To make contact with the bosonic Kondo model, it is useful to split
the first term into a real and an imaginary part, so that
\bea
\frac{- i \Delta}{\omega- \xi}= \Delta A_f( \omega) - i \Delta  { \rm Re}\left( \frac{1}{\om - \xi}\right)
\eea
One can then move the second term above into the logarithm,
writing
\bea
\frac{-i \Delta}{q \pi} { \rm Re}\left( \frac{1}{\om - \xi}\right)
{\ln }\left(\frac{\xi^*}{\omega}\right)- i \Delta
 { \rm Re}\left( \frac{1}{\om - \xi}\right)
= \frac{-i \Delta}{q \pi} { \rm Re}\left( \frac{1}{\om - \xi}\right)
\left[\ln { T_K \over \om}+i \pi \tilde n_b 
\right]
\eea
 to obtain
\bea
P_o( \omega- i \delta)
= A_f( \omega) \left[ \Delta
- \frac{ \omega - 2 \zeta}{\pi q} 
\left(\ln { T_K  \over \om}+ {i \pi \tilde n_b}\right)
\right]
\eea


\begin{references}



\bibitem{mathur}N. D. Mathur ND, F. M. Grosche , S. R. Julian , I. R.  Walker,
, D. M. Freye , R. K. W. Haselwimmer  and  G. G. Lonzarich ,   Nature, {\bf 394}, 39 (1998).

\bibitem{grosche}F. M. Grosche, P. Agarwal, S. R. Julian, N. J. Wilson, R. K. W. Haselwimmer, S. J. S. Lister, N. D. Mathur, F. V. Carter, S. S. Saxena, G. G. Lonzarich,
cond-mat/9812133 to be published (1999).

\bibitem{lohneyson}
H. von L\"ohneysen,  Journal of Physics-Cond. Matt. {\bf 8}
9689 (1996). 

\bibitem{schroeder}A. Schr\"oder, G. Aeppli, E.Bucher, R. Ramazashvili , P. Coleman,  Physical Review Letters {\bf 80}, 5623 (1998).

\bibitem{stewart}K. Heuser , E. W. Scheidt, T. Schreiner , G. R. Stewart , 
Physical Review B {\bf 57}, no.8, 4198 (1998).

\bibitem{hilbert2}O. Stockert O, H. von L\"ohneysen H, A. Rosch A, N. Pyka , M. Loewenhaupt, 
Phys. Rev. Lett. {\bf 80},
5627, (1998),

\bibitem{steglich}
F. Steglich, P. Gegenwart, R. Helfrich, C. Langhammer, P. Hellmann, L. Donnevert, C. Geibel, M. Lang,  G. Sparn,
W. Assmus, G. R. Stewart and A. Ochiai, 
Zeitschrift fur Physik B-Condensed Matter, {\bf 103}, 235 (1997).



\bibitem{hertz}John Hertz,  Phys. Rev. B {\bf 14} 525, (1976).

\bibitem{millisqcp}A. J. Millis, 
Phys. Rev. B {\bf 48}, 7183 (1993). 


\bibitem{rosch99}A. Rosch,  
Phys. Rev. Lett. {\bf 82}, 4280 (1999).




\bibitem{sces}P. Coleman,  
Physica B {\bf 259-261}, 353 (1999). 
 

\bibitem{schroeder99}A. Schroder et al, ``Spin Condensation'', submittted
for publication (1999). 

\bibitem{qimiao99}Qimiao Si, J. Lleweilun Smith (Rice U.), Kevin Ingersent,
 to be published, 
cond-mat/9905006 (1999).



 
\bibitem{doniach77}  S. Doniach, {\sl Physica B \& C} {\bf 91}, 231, (1977).

\bibitem{coleman} P. Coleman, Phys. Rev. B {\bf 29}, 3035 (1984).

\bibitem{read}N. Read, D. M. Newns  and S.  Doniach,
Phys. Rev. B {\bf 30}, 3841 (1984).

\bibitem{readimp}N. Read, D.M. Newns, J. Phys. C {\bf 16}, 3273 (1968).

\bibitem{auerbach}A. Auerbach and K. Levin, 
Phys. Rev. Lett. {\bf 57}877 (1986).

\bibitem{millis}A. J. Millis and P. A. Lee,  Physical
Review B-Condensed Matter, {\bf 35}, 3394 (1987).

\bibitem{auerbach88}A. Auerbach and D. P.  Arovas, 
Phys. Rev. Lett. {\bf 61}, .615-20 (1988).


\bibitem{arovas}D. P. Arovas D. P.  and A. 
Auerbach,
 Phys. Rev {\bf B 38}, 316-211, 1 July 1988.

\bibitem{abrikosov}A. A. Abrikosov,    Physics {\bf 2}, 5 (1965).

\bibitem{grptheory}For a reference on Young tableaux, see e.g.
M. Hammermesh, ``Group Theory and its Application to Physical Problems'',
pp 198,  Addison  Wesley, (1962) or H. F. Jones, ``Groups Representations
and Physics, Institute of Physics, (1990).

\bibitem{georges}O. Parcollet and  A. Georges,  Phys. Rev. Lett. { \bf 79}, 4665 (1997).

\bibitem{zarand}A. Jerez A, N. Andrei N and G.  Zarand
Phys. Rev.  B, {\bf 58} 3814 (1998).

\bibitem{okubo}S. Okubo, J. Math. Phys {\bf 18}, 2382 (1977).

\bibitem{gan}J. Gan , P. Coleman, N. Andrei   Phys. Rev. Lett. {\bf 68}, 3476, (1992); J. Gan and P. Coleman, Physica B, {\bf 171}, 3 (1991).


\bibitem{pepin}C. P\'epin and M. Lavagna ,  Phys. Rev. B, {\bf 59}, no.19,12180 (1999);Z. Phys. B {\bf 103}, 259 (1997).

\bibitem{ngai}T. K. Ng and C. H. Cheng, cond-mat/9802080 (1998).

\bibitem{supergroups}I. Bars, 
Physica D, {\bf 15D}, 42 (1985); I. Bars, Lectures in Applied Mathematics,(American Mathematical Society)  {\bf 21}, 17, (1985). 

\bibitem{blandin} P.Nozi\`eres and A. Blandin, J. Phys. (Paris) {\bf 41} 193 (1980).

\bibitem{pepin2000}C. P\'epin, P. Coleman and A. M. Tsvelik, to be
published.


 
\end{references}
\end{document}